\documentclass[floatfix,twocolumn,showpacs,preprintnumbers,amsmath,amssymb,pra,superscriptaddress,longbibliography]{revtex4-1}
\usepackage{color}
\usepackage[usenames,dvipsnames,svgnames,table]{xcolor}
\usepackage[colorlinks=true,linkcolor=blue,urlcolor=blue,citecolor=blue]{hyperref}
\usepackage{mathtools}
\usepackage{graphicx}
\usepackage{dcolumn}
\usepackage{array}
\usepackage{lipsum}
\usepackage{bm}
\usepackage{subfigure}
\usepackage{amssymb}
\usepackage{multirow}
\usepackage{tabularx}
\usepackage{amsmath}
\usepackage{braket}
\graphicspath{{plots/}}
 \usepackage{lipsum}
\usepackage{mathrsfs}

\newcommand{\beq}{\begin{equation}}
\newcommand{\eeq}{\end{equation}}
\newcommand{\bea}{\begin{eqnarray}}
\newcommand{\eea}{\end{eqnarray}}

\newcommand{\qv}{{\bf q}}

\renewcommand{\vec}[1]{\mathbf{#1}}

\begin{document}
\title{
Nonlinear density response from imaginary-time correlation functions:\\ \emph{Ab initio} path integral Monte Carlo simulations of the warm dense electron gas
}

\author{Tobias Dornheim}
\email{t.dornheim@hzdr.de}

\affiliation{Center for Advanced Systems Understanding (CASUS), D-02826 G\"orlitz, Germany}
\affiliation{Helmholtz-Zentrum Dresden-Rossendorf (HZDR), D-01328 Dresden, Germany}

\author{Zhandos A. Moldabekov}

\affiliation{Center for Advanced Systems Understanding (CASUS), D-02826 G\"orlitz, Germany}
\affiliation{Helmholtz-Zentrum Dresden-Rossendorf (HZDR), D-01328 Dresden, Germany}

\author{Jan Vorberger}
\affiliation{Helmholtz-Zentrum Dresden-Rossendorf (HZDR), D-01328 Dresden, Germany}

\begin{abstract}
The \emph{ab initio} path integral Monte Carlo (PIMC) approach is one of the most successful methods in quantum many-body theory. A particular strength of this method is its straightforward access to imaginary-time correlation functions (ITCF). For example, the well-known density--density ITCF $F(\mathbf{q},\tau)$ allows one to estimate the linear response of a given system for all wave vectors $\mathbf{q}$ from a single simulation of the unperturbed system. Moreover, it constitutes the basis for the reconstruction of the dynamic structure factor $S(\mathbf{q},\omega)$---a key quantity in state-of-the-art scattering experiments.
In this work, we present analogous relations between the nonlinear density response in quadratic and cubic order of the perturbation strength and generalized ITCFs measuring correlations between up to four imaginary-time arguments. As a practical demonstration of our new approach, we carry out simulations of the warm dense electron gas and find excellent agreement with previous PIMC results that had been obtained with substantially larger computational effort. In addition, we give a relation between a cubic ITCF and the triple dynamic structure factor $S(\mathbf{q}_1,\omega_1;\mathbf{q}_2,\omega_2)$, which evokes the enticing possibility to study dynamic three-body effects on an \emph{ab initio} level.
\end{abstract}

\maketitle

\section{Introduction\label{sec:introduction}}

Having originally been introduced for the description of $^4$He in the 1960s~\cite{Fosdick_PhysRev_1966,Jordan_PhysRev_1968}, the \emph{ab initio} path integral Monte Carlo (PIMC) method~\cite{Berne_JCP_1982,Takahashi_Imada_PIMC_1984,cep} has emerged as one of the most successful methods in statistical physics and related disciplines. In particular, the PIMC method allows for the in principle exact solution of the full quantum many-body problem in thermodynamic equilibrium and has been pivotal for our understanding of important physical effects such as superfluidity~\cite{cep,ultracold2,Kwon_PRL_1999}, Bose-Einstein condensation~\cite{Saito_Japan_2016,Pilati_2010}, and the crystallization of quantum systems~\cite{PhysRevLett.86.870,PhysRevLett.76.4572,PhysRevLett.86.3851,PhysRevLett.95.235006,PhysRevB.84.075130}. Moreover, sophisticated sampling techniques~\cite{boninsegni1,boninsegni2} allow for the efficient simulation of up to $N\sim10^4$ bosons or boltzmannons (i.e., hypothetical distinguishable quantum particles). Consequently, the PIMC method has been successfully applied to a gamut of physical systems such as ultracold atoms~\cite{Boninsegni1996,Filinov_PRA_2012,Filinov_PRA_2016,Boninsegni_MaxEnt_revisited_2018}, exotic supersolids~\cite{Kora2019,Saccani_Supersolid_PRL_2012,RevModPhys.84.759}, and confined nano-clusters~\cite{Draeger_PRL_2003,mezza,Filinov_PRB_2008,Boninsegni_PRA_2013,Dornheim_PRB_2015}. While the PIMC simulation of quantum degenerate Fermi systems is rendered substantially more involved by the notorious fermion sign problem~\cite{troyer,dornheim_sign_problem}, the last decade has witnesses a spark of activity in this direction~\cite{dornheim_POP,dornheim_jcp,Yilmaz_JCP_2020,lee2020phaseless,Malone_JCP_2015,Malone_PRL_2016,Dornheim_NJP_2015,Schoof_PRL_2015} as well, most notably in the context of so-called warm dense matter~\cite{Driver_Militzer_PRL_2012,Brown_PRL_2013,Militzer_Driver_PRL_2015,dornheim_prl,groth_prl,review}.

A particular strength of the PIMC method is its straightforward access to imaginary-time correlation functions (ITCF)~\cite{Berne_JCP_1983} such as the imaginary-time version of the intermediate scattering function 
\begin{eqnarray}\label{eq:F}
 F(\mathbf{q},\tau) = \braket{\tilde n(\mathbf{q},0) \tilde n(\mathbf{-q},\tau)}\ ,
 \end{eqnarray}
with $\tilde n(\mathbf{q},\tau)$ being the density operator in reciprocal space evaluated at imaginary times $\tau\in[0,\beta]$~\cite{beta_note}. First and foremost, Eq.~(\ref{eq:F}) gives one direct access to the linear density response function~\cite{quantum_theory} in the static limit
\begin{eqnarray}\label{eq:LRT}
\chi(\mathbf{q}) = - \frac{1}{V} \int_0^\beta d\tau\ F(\mathbf{q},\tau)\ .
\end{eqnarray}
Indeed, the importance of Eq.~(\ref{eq:LRT}) can hardly be overstated, as it allows one to obtain the complete wave-vector dependence of the density response function $\chi(\mathbf{q})$ from a single simulation of the unperturbed system of interest, see, e.g., Refs.~\cite{PhysRevLett.105.070401,dornheim_ML,dynamic_folgepaper} for recent applications.
In addition, $F(\mathbf{q},\tau)$ is connected to the dynamic structure factor $S(\mathbf{q},\omega)$ via the relation
\begin{eqnarray}\label{eq:S_from_F}
F(\mathbf{q},\tau) = \int_{-\infty}^\infty d\omega\ S(\mathbf{q},\omega) e^{-\omega\tau}\ .
\end{eqnarray}
While the numerical inversion of Eq.~(\ref{eq:S_from_F}) constitutes a notoriously hard problem~\cite{Jarrell_Gubernatis_PhysRep_1996}, it nevertheless offers the enticing possibility to obtain dynamic properties that can be directly compared to experimental measurements~\cite{siegfried_review} without any approximations of exchange--correlation effects, see Refs.~\cite{Vitali_PRB_2010,Saccani_Supersolid_PRL_2012,Filinov_PRA_2012,dornheim_dynamic} for a selection of recent applications. Other examples of relevant ITCFs include the well-known Matsubara Green function~\cite{boninsegni1} and the velocity--velocity CF~\cite{Rabani_PNAS_2002}.

Unfortunately, the relations (\ref{eq:F})-(\ref{eq:S_from_F}) have hitherto been limited to the linear response regime, i.e., the limit of a weak perturbation whose effect can be described in first-order of the perturbation strength. This is highly unsatisfactory as nonlinear effects are known to play an important role in many situations, such as the excitation spectrum of graphene~\cite{PhysRevLett.105.097401,Mikhailov_PRL,Cox2016}, the optical excitation and ionization of atoms~\cite{keldysh,schafer_prl_93}, and state-of-the-art experiments with warm dense matter~\cite{Dornheim_PRL_2020,Dornheim_PRR_2021}. Therefore, the first PIMC results for the nonlinear density response have been obtained by actually applying an external perturbation to the system and subsequently measuring its response. Obviously, this procedure is exceedingly demanding as it requires to perform $N_A\sim10$ independent PIMC simulations to compute the nonlinear response \emph{for a single wave vector $\mathbf{q}$}. At the same time, we note that nonlinear effects are substantially more sensitive to exchange--correlation effects~\cite{Dornheim_PRR_2021}, which makes exact PIMC simulations all the more important.

In the present work, we overcome this limitation by presenting analogous relations between density--density ITCFs depending on two and three $\tau$-arguments and the generalized quadratic and cubic static response functions $\mathscr{Y}(\qv_1,\qv_2)$ and $\mathscr{Z}(\qv_1,\qv_2,\qv_3)$, respectively. As a practical example of high relevance, we carry out simulations of the uniform electron gas (UEG) at warm dense matter conditions and compare the results from our new scheme to previous PIMC data~\cite{Dornheim_PRL_2020,Dornheim_PRR_2021} that have been obtained from the perturbed electrons gas with substantially larger computational effort.
In addition, we outline a strategy for an analytic continuation of these new ITCFs, which, in principle, gives one access to generalized dynamic structure factors involving the time-dependent correlations between three and four particles.

The paper is organized as follows: In Sec.~\ref{sec:theory}, we introduce the required theoretical background, including the UEG (\ref{sec:UEG}), the PIMC method and the related estimation of ITCFs (\ref{sec:PIMC}) and the theory of the nonlinear density response (\ref{sec:nonlinear}). In Secs.~\ref{sec:quadratic_theory} and \ref{sec:cubic_theory}, we present the new relations between different ITCFs and the generalized density response on a quadratic and cubic level, respectively. This is followed by a brief discussion of the relation between the triple dynamic structure factor $S(\mathbf{q}_1,\omega_1;\mathbf{q}_2,\omega_2)$ and the imaginary-time structure of the system in Sec.~\ref{sec:analytic}. Sec.~\ref{sec:results} is devoted to the presentation of our new simulation results for the warm dense UEG and is divided into a discussion of the quadratic response (\ref{sec:quadratic_results}) and the cubic response (\ref{sec:cubic_results}). The paper is concluded by a concise summary and outlook in Sec.~\ref{sec:summary}.

\section{Theory\label{sec:theory}}

\subsection{Uniform electron gas\label{sec:UEG}}

Throughout this work, we restrict ourselves to a fully unpolarized electron gas (i.e., $N^\uparrow=N^\downarrow=N/2$, with $N^\uparrow$ and $N^\downarrow$ being the number of spin-up and spin-down electrons) in a cubic simulation cell of volume $V=L^3$ at a constant temperature $T$.
Moreover, all formulae and results are given in Hartree atomic units (i.e., $\hbar=m_e=k_\textnormal{B}=1$), which leads to the UEG Hamiltonian~\cite{review}
\begin{eqnarray}\label{eq:Hamiltonian_UEG}
\hat H_\textnormal{UEG} = -\frac{1}{2}\sum_{l=1}^N\nabla_l^2 + \frac{1}{2}\sum_{l\neq p}^N\phi_\textnormal{E}(\hat{\mathbf{r}}_l,\hat{\mathbf{r}}_p)\ ,
\end{eqnarray}
where $\phi_\textnormal{E}(\hat{\mathbf{r}}_l,\hat{\mathbf{r}}_p)$ is the Ewald pair potential as defined e.g. in Ref.~\cite{Fraser_Foulkes_PRB_1996}.

In addition, we mention that the UEG in the warm dense matter regime is typically defined by two characteristic parameters~\cite{Ott2018}: i) the density parameter (also known as Wigner-Seitz radius) $r_s=\overline{r}/a_\textnormal{B}$, where $\overline{r}$ and $a_\textnormal{B}$ are the average particle distance and first Bohr radius, and ii) the degeneracy temperature $\theta=T/E_\textnormal{F}$, with $E_\textnormal{F}$ being the usual Fermi energy~\cite{quantum_theory}. More specifically, $r_s$ plays the role of a \emph{quantum coupling parameter}, so that one recovers the ideal Fermi gas in the limit of $r_s\to0$, whereas the UEG is known to enter the strongly coupled electron liquid regime for $r_s\gtrsim10$~\cite{dornheim_electron_liquid}, and eventually forms a Wigner crystal for $r_s\gtrsim100$~\cite{Drummond_Wigner_2004}. The second parameter, $\theta$, gives direct information about the importance of quantum degeneracy effects, with $\theta\gg 1$ ($\theta\ll 1$) indicating a fully classical (fully quantum degenerate) system. The warm dense matter regime is defined by the condition  $r_s\sim\theta\sim 1$, which leads to a remarkable interplay of Coulomb coupling, quantum degeneracy, and thermal excitation effects~\cite{wdm_book,new_POP}.

While the UEG is employed in the present study as a convenient model system to demonstrate and benchmark the new ITCFs defined below, we mention that it is one of the most fundamental model systems in physics, quantum chemistry, and related disciplines~\cite{loos}, and has been pivotal for the development of important physical concepts. Furthermore, the accurate description of the UEG based on \emph{ab inito} quantum Monte Carlo methods~\cite{Foulkes_RevModPhys_2001} constitutes an important end in itself and has been of direct use for many applications, most notably the development of exchange--correlation functionals for density functional theory~\cite{Jones_RevModPhys_2015}.

\subsection{Path integral Monte Carlo\label{sec:PIMC} and ITCFs}

The UEG as it has been introduced in the previous section is completely described by the canonical partition function
\begin{eqnarray}\label{eq:Z}
Z_{\beta,N,V} &=& \frac{1}{N^\uparrow! N^\downarrow!} \sum_{\sigma^\uparrow\in S_N} \sum_{\sigma^\downarrow\in S_N} \textnormal{sgn}(\sigma^\uparrow,\sigma^\downarrow)\\\nonumber & & \times \int d\mathbf{R} \bra{\mathbf{R}} e^{-\beta\hat H} \ket{\hat{\pi}_{\sigma^\uparrow}\hat{\pi}_{\sigma^\downarrow}\mathbf{R}}\ ,
\end{eqnarray}
with $\mathbf{R}=(\mathbf{r}_1,\dots,\mathbf{r}_N)^T$ containing the coordinates of all $N$ particles,  and $\hat{\pi}_{\sigma^\uparrow}$ ($\hat{\pi}_{\sigma^\downarrow}$) being the permutation operator corresponding to a particular element $\sigma^\uparrow$ ($\sigma^\downarrow$) from the permutation group $S_N$. Furthermore, $\textnormal{sgn}(\sigma^\uparrow,\sigma^\downarrow)$ denotes the sign function, which is positive (negative) for an even (odd) number of pair permutations~\cite{Dornheim_permutation_cycles}.

The problem with Eq.~(\ref{eq:Z}) is that a direct evaluation of the matrix elements of the density operator $\hat\rho=e^{-\beta\hat H}$ is not possible as the kinetic and interaction parts of the Hamiltonian Eq.~(\ref{eq:Hamiltonian_UEG}) do not commute. The basic idea behind the PIMC method~\cite{cep} is to use an exact (semi-)group property of the density matrix, which eventually leads to a modified expression for $Z_{\beta,N,V}$,
\begin{eqnarray}\label{eq:Z_modified}
Z_{\beta,N,V} &=& \frac{1}{N^\uparrow! N^\downarrow!} \sum_{\sigma^\uparrow\in S_N} \sum_{\sigma^\downarrow\in S_N} \textnormal{sgn}(\sigma^\uparrow,\sigma^\downarrow)\\\nonumber & & \times \int d\mathbf{R}_0\dots d\mathbf{R}_{P-1}
\bra{\mathbf{R}_0}e^{-\epsilon\hat H}\ket{\mathbf{R}_0}\\\nonumber & & \times \bra{\mathbf{R}_1}e^{-\epsilon\hat H}\ket{\mathbf{R}_1} \dots 
\bra{\mathbf{R}_{P-1}} e^{-\beta\hat H} \ket{\hat{\pi}_{\sigma^\uparrow}\hat{\pi}_{\sigma^\downarrow}\mathbf{R}_0}\ ,
\end{eqnarray}
with the definition $\epsilon=\beta/P$.
Comparing Eqs.~(\ref{eq:Z}) and (\ref{eq:Z_modified}), we see that the partition function has now been expressed as the integral over $P$ density matrices, with each one being evaluated at a $P$-times higher temperature.

\begin{figure}\centering\includegraphics[width=0.485\textwidth]{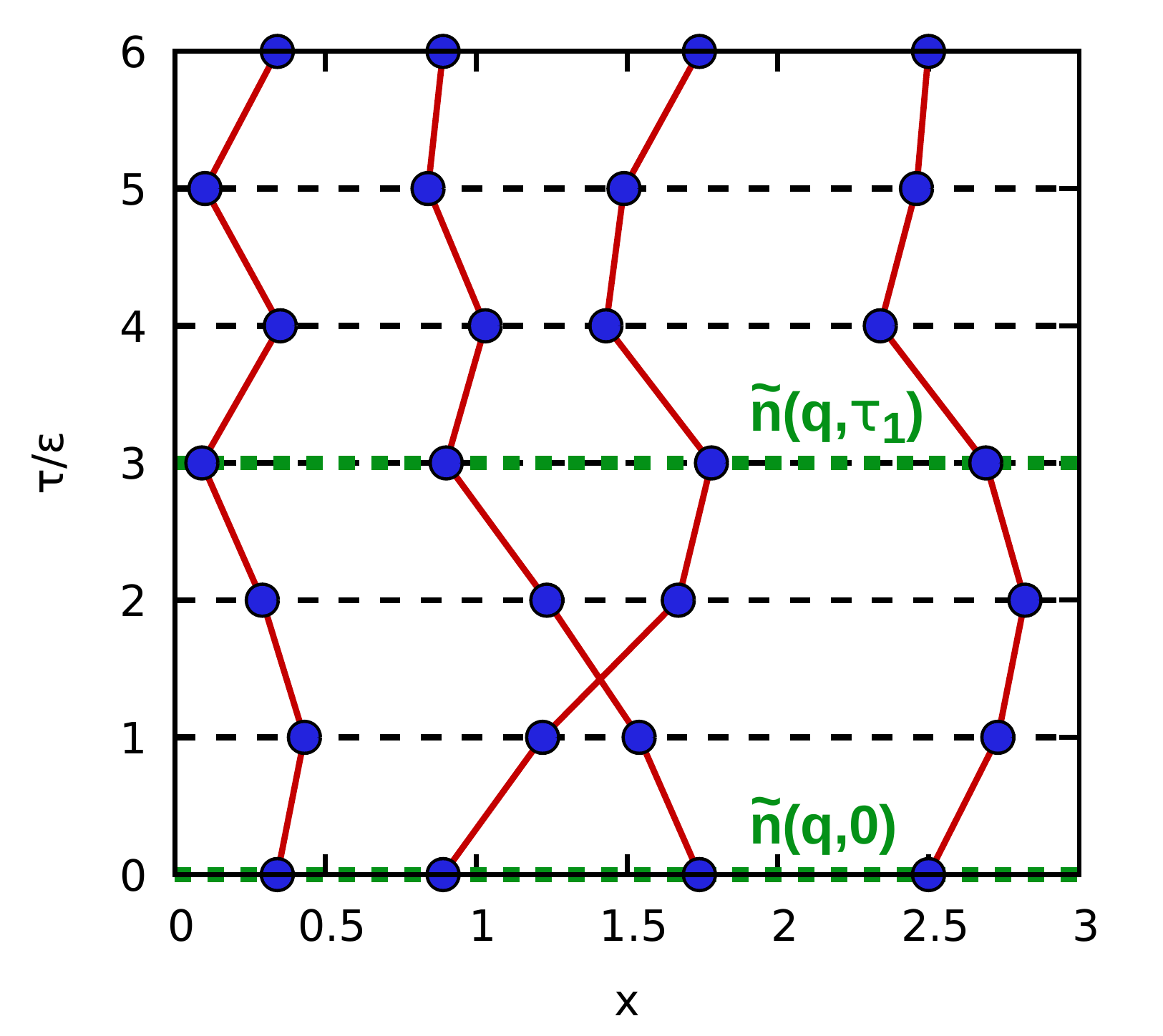}
\caption{\label{fig:ITCF} Illustration of the PIMC method and the estimation of the ITCF $F(\mathbf{q},\tau)$ [cf.~Eq.~(\ref{eq:F})]. Shown is a configuration of $N=4$ particles in the $x$-$\tau$-plane. Each particle is represented by a \emph{closed} path in the imaginary time $\tau\in[0,\beta]$, which a single pair exchange being present around the center. The green horizontal lines illustrate the evaluation of the density operator in reciprocal space $\tilde n(\mathbf{q},\tau)$ [cf.~Eq.~(\ref{eq:n_tilde})] on two different time slices, giving direct access to $F(\mathbf{q},\tau)$.
}
\end{figure}

This is illustrated in Fig.~\ref{fig:ITCF}, where we show a configuration of $N=4$ particles in the $\tau$-$x$-plane with $P=6$. Evidently, all particles are now represented by entire paths over distinct sets of coordinates in the imaginary time $\tau$, as each high-temperature factor in Eq.~(\ref{eq:Z_modified}) can be interpreted as a propagation by a time-step $\epsilon$. Furthermore, we note that that the coordinates at $\tau=0$ and $\tau=\beta$ are equal, i.e., $\mathbf{R}_0=\mathbf{R}_{P}$, which is a direct consequence of the definition of $Z$ as the sum over diagonal elements, c.f.~Eq.~(\ref{eq:Z}). Furthermore, the configuration depicted in Fig.~\ref{fig:ITCF} contains a single pair permutation in the middle, which results in an exchange-cycle containing two particles~\cite{Dornheim_permutation_cycles}. Consequently, the sign function in Eqs.~(\ref{eq:Z}) and (\ref{eq:Z_modified}) is negative, which leads to the \emph{fermion sign problem} discussed below.

From a practical point of view, the important achievement of Eq.~(\ref{eq:Z_modified}) is that each of the involved  density matrices can be reasonably approximated by a suitable high-temperature approximation. For example, even the primitive factorization $e^{-\epsilon\hat H}\approx e^{-\epsilon\hat V}e^{-\epsilon\hat K}$ becomes exact in the limit of large $P$, which is ensured by the well-known Trotter formula~\cite{trotter}. For completeness, we mention that more sophisticated factorizations of $\hat\rho$ allow for the same level of accuracy with a more favourable scaling with respect to $P$~\cite{sakkos_JCP_2009,brualla_JCP_2004}. Yet, the primitive factorization actually constitutes the preferred option in the context of the present work, as it directly determines the number of discrete $\tau$-points on which any ITCF can be evaluated.

For example, the PIMC estimator for the IT intermediate scattering function from Eq.~(\ref{eq:F}) is given by
\begin{eqnarray}\label{eq:F_estimator}
F(\mathbf{q},\tau_j) &=& \frac{1}{Z_{\beta,N,V}} \frac{1}{N^\uparrow! N^\downarrow!} \sum_{\sigma^\uparrow\in S_N} \sum_{\sigma^\downarrow\in S_N} \textnormal{sgn}(\sigma^\uparrow,\sigma^\downarrow)\\\nonumber & & \times \int d\mathbf{R}_0\dots d\mathbf{R}_{P-1}
\bra{\mathbf{R}_0}e^{-\epsilon\hat H}\tilde{n}(\mathbf{q})\ket{\mathbf{R}_0}\\\nonumber & & \times \bra{\mathbf{R}_1}e^{-\epsilon\hat H}\ket{\mathbf{R}_1} \dots \bra{\mathbf{R}_j}e^{-\epsilon\hat H}\tilde{n}(-\mathbf{q})\ket{\mathbf{R}_j} \dots \\ \nonumber & & \times
\bra{\mathbf{R}_{P-1}} e^{-\beta\hat H} \ket{\hat{\pi}_{\sigma^\uparrow}\hat{\pi}_{\sigma^\downarrow}\mathbf{R}_0}\ ,
\end{eqnarray}
and is, thus, available for discrete imaginary times $\tau_j=j\epsilon$, $j=0,\dots,P$, with $F(\mathbf{q},0)=F(\mathbf{q},\beta)$. As the PIMC results for different ITCFs are used as input for a subsequent numerical integration, the accuracy of our results for the different density response functions is primarily determined by the corresponding integration error, and not by the employed factorization of the density matrix. A more vivid illustration of Eq.~(\ref{eq:F_estimator}) is included by the dotted green horizontal lines in Fig.~\ref{fig:ITCF}. More specifically, the evaluation of $F(\mathbf{q},\tau_j)$ requires us to compute the density in reciprocal space on two different imaginary-time slices, and then average over their product.

Let us conclude this section with a brief note on the aforementioned \emph{fermion sign problem}. As the high dimensionality of the path integrals in Eqs.~(\ref{eq:Z_modified}) and (\ref{eq:F_estimator}) renders classical quadrature methods impractical, the PIMC method is based on the Metropolis Monte Carlo algorithm~\cite{metropolis}. Specifically, the configurations of particle coordinates are generated randomly and appear in the resulting Markov chain with a probability that is proportional to their respective contribution to $Z_{\beta,N,V}$. While this is fairly straightforward in the case of bosons and boltzmannons, the fermionic anti-symmetry with respect to particle exchange leads to negative contributions to the total partition function, which, in turn, cannot be interpreted as a probability distribution. While this sampling problem can be circumvented in practice, it nevertheless leads to a cancellation of positive and negative terms in the computation of any expectation value. In fact, it is well known that this effect leads to an exponential increase in the computation time when either the system size $N$ or inverse temperature $\beta$ are increased; see Ref.~\cite{dornheim_sign_problem} for an accessible review of this issue. Therefore, the fermion sign problem constitutes the main limitation of our simulations in the present work.

For completeness, we mention that the restricted PIMC method, which formally avoids the sign problem by introducing the uncontrolled \emph{fixed-node approximation}~\cite{Ceperley1991}, cannot be used here, as it breaks the $\tau$-symmetry in the path-integral picture and, therefore, does not have the direct access to ITCFs of the direct PIMC method.

\subsection{Nonlinear density response theory\label{sec:nonlinear}}

In both linear and nonlinear response theory alike, we are interested in the response of the original system (described by the unperturbed Hamiltonian $\hat H_0$) to an external harmonic perturbation~\cite{moroni,moroni2,bowen2,dornheim_pre,groth_jcp,Dornheim_PRL_2020},
\begin{eqnarray}\label{eq:Hamiltonian_perturbed}
\hat H = \hat H_0 + 2A\sum_{l=1}^N \textnormal{cos}\left(
\mathbf{q}\cdot\hat{\mathbf{r}}_l
\right)\ ,
\end{eqnarray}
with $\mathbf{q}$ being the corresponding wave vector and $A$ the perturbation amplitude. For sufficiently small values of $A$, the resulting density profile is described accurately by the linear response function $\chi(\mathbf{q})$ introduced in Eq.~(\ref{eq:LRT}) above, which gives
\begin{eqnarray}\label{eq:density_LRT}
n_\textnormal{LRT}(\mathbf{r}) = n_0 + 2A\textnormal{cos}\left(
\mathbf{q}\cdot\mathbf{r}
\right)\chi(\mathbf{q})\ ,
\end{eqnarray}
where $n_0$ is the average density. In other words, the density profile simply follows the external perturbation within LRT. Obviously, Eq.~(\ref{eq:density_LRT}) must break down eventually for large $A$ as it would result in \emph{negative} values of $n(\mathbf{r})$ in this case, which is a physical impossibility. Therefore, it has to be replaced by a complete expansion over the integer harmonics of the original perturbation~\cite{Dornheim_PRR_2021}
\begin{eqnarray}
n(\mathbf{r}) = n_0 + 2\sum_{\eta=1}^\infty \left\{ \frac{\braket{\tilde{n}(\eta\mathbf{q})}_{q,A}}{V}\ \textnormal{cos}\left(
\eta\mathbf{q}\cdot\mathbf{r}
\right)\right\}\ ,
\end{eqnarray}
where the notation $\braket{\dots}_{q,A}$ indicates the evaluation of the expectation value with respect to the Hamiltonian of the perturbed system, Eq.~(\ref{eq:Hamiltonian_perturbed}).
The corresponding coefficients can be readily expanded in powers of $A$, and a truncation at the cubic level gives for the first three harmonics
\begin{eqnarray}
\frac{\braket{\tilde{n}(\mathbf{q})}_{q,A}}{V} &=& \chi(\mathbf{q})A + \chi^{(1,\textnormal{cubic})}(\mathbf{q})A^3 + \dots \label{eq:n_first} \\
\frac{\braket{\tilde{n}(2\mathbf{q})}_{q,A}}{V} &=& \chi^{(2)}(\mathbf{q}) A^2 + \dots\label{eq:n_second}\\
\frac{\braket{\tilde{n}(3\mathbf{q})}_{q,A}}{V} &=& \chi^{(3)}(\mathbf{q}) A^3 + \dots\label{eq:n_third}\ ,
\end{eqnarray}
where the additional coefficients are, by definition, given by the nonlinear density response functions $\chi^{(1,\textnormal{cubic})}(\mathbf{q})$, $\chi^{(2)}(\mathbf{q})$, and $\chi^{(3)}(\mathbf{q})$ in which we are interested in the present work.

In fact, Eqs.~(\ref{eq:n_first})-(\ref{eq:n_third}) have been recently used~\cite{Dornheim_PRL_2020,Dornheim_PRR_2021} to compute these functions for the UEG by fitting their RHS. to the PIMC data for density in reciprocal space at the different harmonics on the LHS. While being in principle exact, this procedure requires one to carry out independent PIMC simulations for $N_A\sim\mathcal{O}(10)$ values of $A$ \emph{for each individual wave vector $\mathbf{q}$}. Therefore, obtaining the full wave-number dependence of the nonlinear density response for a single density-temperature combination requires $N_{r_s,\theta}\sim\mathcal{O}(10^2)$ PIMC simulations of Eq.~(\ref{eq:Hamiltonian_perturbed}), which quickly becomes infeasible and certainly rules out extensive parameter scans as they have been presented for the linear response on the basis of Eq.~(\ref{eq:LRT}).

\subsection{Quadratic density response\label{sec:quadratic_theory}}
The generalized quadratic density response function (see Appendix~\ref{sec:derivations} for details) is connected to the imaginary-time structure of the system by the relation
\begin{eqnarray}\label{eq:Y_imaginary}
\lefteqn{\mathscr{Y}(\qv_1,\qv_2)=\frac{1}{2L^3}\int\limits_0^{\beta}d\tau_1\int\limits_0^{\beta}d\tau_2}&&\\\nonumber
&&\times\langle
\tilde n(\qv_1+\qv_2,0)\tilde n(-\qv_1,-\tau_1)\tilde n(-\qv_2,-\tau_2)
\rangle\,.
\end{eqnarray}
Note that we define 
\begin{eqnarray}\label{eq:n_tilde}
\tilde n(\mathbf{q},\tau) = \sum_{l=1}^N \textnormal{exp}\left(
-i\mathbf{q}\cdot\hat{\mathbf{r}}_{l,\tau}
\right)
\end{eqnarray}
to be not normalized, and $\mathbf{r}_{l,\tau}$ denotes the coordinates of particle $l$ at an imaginary time $\tau$.

While $\mathscr{Y}(\qv_1,\qv_2)$ contains the complete information about the quadratic density response, its particular relation to the (static) quadratic response at the second harmonic $\chi^{(2)}(\mathbf{q})$ is given by~\cite{Dornheim_PRR_2021} 
\begin{eqnarray}
\chi^{(2)}(\mathbf{q})&=\mathscr{Y}(\vec k-\vec q,\vec q)\delta_{\vec k,2\vec q}, \label{eq:chi2_Y}
\end{eqnarray}
which, when being combined with Eq.~(\ref{eq:Y_imaginary}), leads to
\begin{eqnarray}\label{eq:chi2_int}
\chi^{(2)}(\mathbf{q}) = {\frac{1}{2L^3}} \int\limits_0^{\beta}d\tau_1\int\limits_0^{\beta}d\tau_2\ F^{(2)}(\mathbf{q},\tau_1,\tau_2)\ ,
\end{eqnarray}
with the definition of the quadratic ITCF
\begin{eqnarray}\label{eq:F2}
F^{(2)}(\mathbf{q},\tau_1,\tau_2) = \braket{\tilde n(2\mathbf{q},0)\tilde n(-\mathbf{q},\tau_1)\tilde n(-\mathbf{q},\tau_2)}\ .
\end{eqnarray}
For the purpose of our analysis, we also find it useful to define an integrated quadratic ITCF as
\begin{eqnarray}\label{eq:I2}
I^{(2)}(\mathbf{q},\tau_1) = \int\limits_0^{\beta}d\tau_2\ F^{(2)}(\mathbf{q},\tau_1,\tau_2)\ ,
\end{eqnarray}
where the dependence on the second imaginary-time argument $\tau_2$ has been integrated out. Conversely, this allows us to re-write Eq.~(\ref{eq:chi2_int}) as
\begin{eqnarray}
\chi^{(2)}(\mathbf{q}) = {\frac{1}{2L^3}} \int\limits_0^{\beta}d\tau_1\ I^{(2)}(\mathbf{q},\tau_1)\ .
\end{eqnarray}

\subsection{Cubic density response\label{sec:cubic_theory}}

The complete information about the general cubic density response function $\mathscr{Z}(\mathbf{q}_1,\mathbf{q}_2,\mathbf{q}_3)$ can be obtained from density-correlations in the imaginary time via the relation 
\begin{eqnarray}\label{eq:Z_imaginary}
\lefteqn{\mathscr{Z}(\mathbf{q}_1,\mathbf{q}_2,\mathbf{q}_3) = \frac{1}{6L^3} \int\limits_0^{\beta}d\tau_1\int\limits_0^{\beta}d\tau_2\int\limits_0^{\beta}d\tau_3}&&\\ \nonumber
&&\times \braket{\tilde n(\mathbf{q}_1+\mathbf{q}_2 +\mathbf{q}_3, 0)\tilde n(-\mathbf{q}_1,\tau_1)\tilde n(-\mathbf{q}_2,\tau_2)\tilde n(-\mathbf{q}_3,\tau_3)}\ .
\end{eqnarray}
The response at the third harmonic of the original perturbation is given by~\cite{Dornheim_PRR_2021}
\begin{eqnarray}
\chi^{(3)}(\mathbf{q}) = \mathscr{Z}(\mathbf{k}-2\mathbf{q},\mathbf{q},\mathbf{q})\delta_{\mathbf{k},3\mathbf{q}}\ ,
\end{eqnarray}
which immediately leads to
\begin{eqnarray}\label{eq:chi3_int}
\chi^{(3)}(\mathbf{q}) = {\frac{1}{6L^3}} \int\limits_0^{\beta}d\tau_1\int\limits_0^{\beta}d\tau_2\int\limits_0^{\beta}d\tau_3 F^{(3)}(\mathbf{q},\tau_1,\tau_2,\tau_3)\ ,\quad\
\end{eqnarray}
with the definition of the cubic ITCF of the third harmonic
\begin{eqnarray}\label{eq:F3}
F^{(3)}(\mathbf{q},\tau_1,\tau_2,\tau_3) &=& \left<\tilde n(3\mathbf{q}, 0)\tilde n(-\mathbf{q},\tau_1)\right.\\\nonumber & &\times\left. \tilde n(-\mathbf{q},\tau_2)\tilde n(-\mathbf{q},\tau_3)\right>\ .
\end{eqnarray}
Similarly to Eq.~(\ref{eq:I2}) given above, we find it useful to introduce the reduced cubic ITCFs
\begin{eqnarray}\label{eq:I3}
I^{(3)}(\mathbf{q},\tau_1,\tau_2) = \int\limits_0^{\beta}d\tau_3\ F^{(3)}(\mathbf{q},\tau_1,\tau_2,\tau_3)\ ,
\end{eqnarray}
and
\begin{eqnarray}\label{eq:J3}
J^{(3)}(\mathbf{q},\tau_1) = \int\limits_0^{\beta}d\tau_2\ I^{(3)}(\mathbf{q},\tau_1,\tau_2)\ .
\end{eqnarray}

\subsection{Relation to higher-order dynamic structure factors\label{sec:analytic}}

The dynamic structure factor (\ref{eq:S_from_F}) can be obtained in various ways in experiments and gives valuable information about short and long range order in the system that can be used to compare theory to measurement~\cite{siegfried_review,falk_wdm}. Higher order correlations enter the dynamic structure factor via dynamic local field corrections~\cite{kugler1} that sum up two-, three-, four-, etc. particle correlations. A more direct analysis of such clusters is possible based on higher order structure factors giving insight into, e.g., bond angle distributions~\cite{Vorberger_c_2020}, lifetimes or correlations of molecules~\cite{ridgeway_2012}, or nonlinear scattering~\cite{Fuchs2015}.

The dynamic \emph{triple} structure factor is given by
\bea
S(\qv_1,\omega_1;\qv_2,\omega_2)&=&\frac{1}{2\pi L^3}\int\limits_{-\infty}^{\infty}\textnormal{d}t_1\int\limits_{-\infty}^{\infty} \textnormal{d}t_2\\\nonumber & &\times
\langle\tilde{n}(\qv_1+\qv_2,0)\tilde{n}(-\qv_1,-t_1)\\\nonumber& &\times\tilde{n}(-\qv_2,-t_2)\rangle
e^{i\omega_1t_1+i\omega_2t_2}
\eea

The inverse transformation in imaginary time is then
\bea\label{eq:triple_reconstruction}
& &\mathscr{F}(\qv_1+\qv_2,0;-\qv_1,\tau_1;-\qv_2,\tau_2)=\\\nonumber& &\int\limits_{-\infty}^{\infty}d\omega_1\int\limits_{-\infty}^{\infty}d\omega_2\,
S(\qv_1,\omega_1;\qv_2,\omega_2)
e^{-\omega_1\tau_1-\omega_2\tau_2}\ ,
\eea
with the definition of the general density--density--density ITCF
\begin{eqnarray}
\mathscr{F}(\mathbf{q}_a,\tau_a;\mathbf{q}_b,\tau_b;\mathbf{q}_c,\tau_c) &=& \braket{\tilde{n}(\mathbf{q}_a,\tau_a)\tilde{n}(\mathbf{q}_b,\tau_b)\tilde{n}(\mathbf{q}_c,\tau_c)}\ . \nonumber \\ & & \label{eq:F2_general}
\end{eqnarray}
Evidently, the quadratic ITCF defined in Eq.~(\ref{eq:F2}) above constitutes a special case of Eq.~(\ref{eq:F2_general}), and the explicit connection is given by
\begin{eqnarray}
F^{(2)}(\mathbf{q},\tau_1,\tau_2)  =  \mathscr{F}(2\mathbf{q},0;-\mathbf{q},\tau_1;-\mathbf{q},\tau_2)\ .
\end{eqnarray}
From a practical point of view, Eq.~(\ref{eq:triple_reconstruction}) can be used as the basis for an analytic continuation~\cite{Jarrell_Gubernatis_PhysRep_1996}, i.e., the numerical inversion to solve for $S(\qv_1,\omega_1;\qv_2,\omega_2)$. 
For each specific combination of wave vectors $\mathbf{q}_1$ and $\mathbf{q}_2$, the LHS of Eq.~(\ref{eq:triple_reconstruction}) is known from a PIMC simulation over the full $\tau_1$- and $\tau_2$-ranges.
The task at hand is then the construction of a suitable trial solution $S_\textnormal{trial}(\qv_1,\omega_1;\qv_2,\omega_2)$ that, when being inserted into Eq.~(\ref{eq:triple_reconstruction}), reproduces the PIMC data for all $(\tau_1,\tau_2)$ combinations. This is, in principle, analogous to the more familiar problem statement given by the reconstruction of the dynamic structure factor $S(\mathbf{q},\omega)$ starting from $F(\mathbf{q},\tau)$, cf.~Eq.~(\ref{eq:S_from_F}) above, for which a gamut of different methods have been presented~\cite{Vitali_PRB_2010,Bertaina_GIFT_2017,Filinov_PRA_2012,Jarrell_Gubernatis_PhysRep_1996,Goulko_PRB_2017}. While the adaption of these approaches to the inversion of Eq.~(\ref{eq:triple_reconstruction}) is conceptually straightforward, the practical implementation of this idea is beyond the scope of the present work and will be pursued in a future publication.

For completeness, we mention that an analogous relation to Eq.~(\ref{eq:triple_reconstruction}) can be found between a \emph{quattro} dynamic structure factor and the general 4-density ITCF that involves the evaluation of four density operators and contains $F^{(3)}(\mathbf{q},\tau_1,\tau_2,\tau_3)$ as a special case.

\section{Results\label{sec:results}}

All results presented in this work have been obtained using a canonical adaption~\cite{mezza} of the worm algorithm introduced by Boninsegni \textit{et al.}~\cite{boninsegni1,boninsegni2}.

\subsection{Quadratic density response\label{sec:quadratic_results}}

\begin{figure*}\centering
\vspace*{-1.5cm}\includegraphics[width=0.485\textwidth]{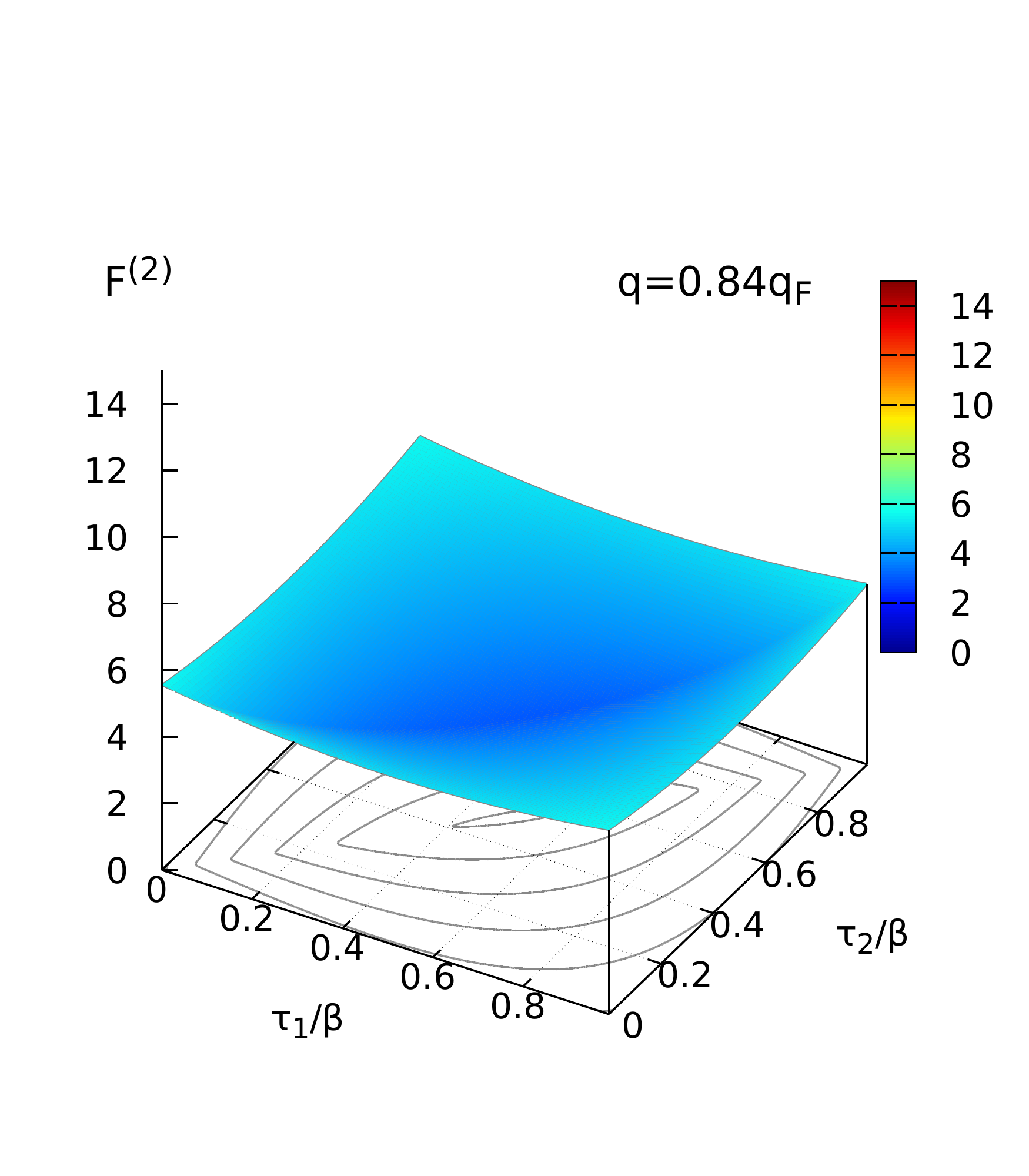}
\includegraphics[width=0.485\textwidth]{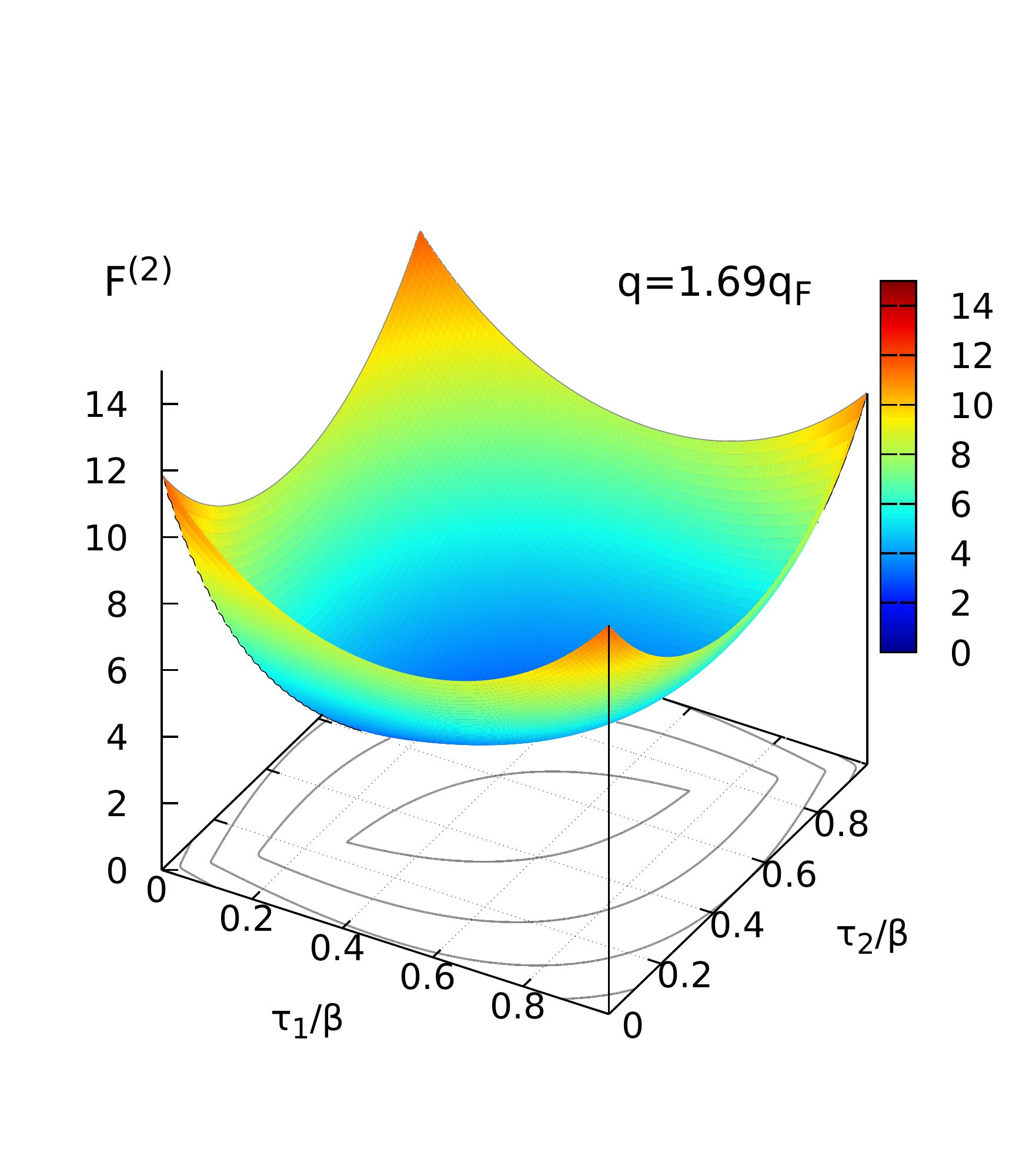}\\\vspace*{-2.5cm}
\includegraphics[width=0.485\textwidth]{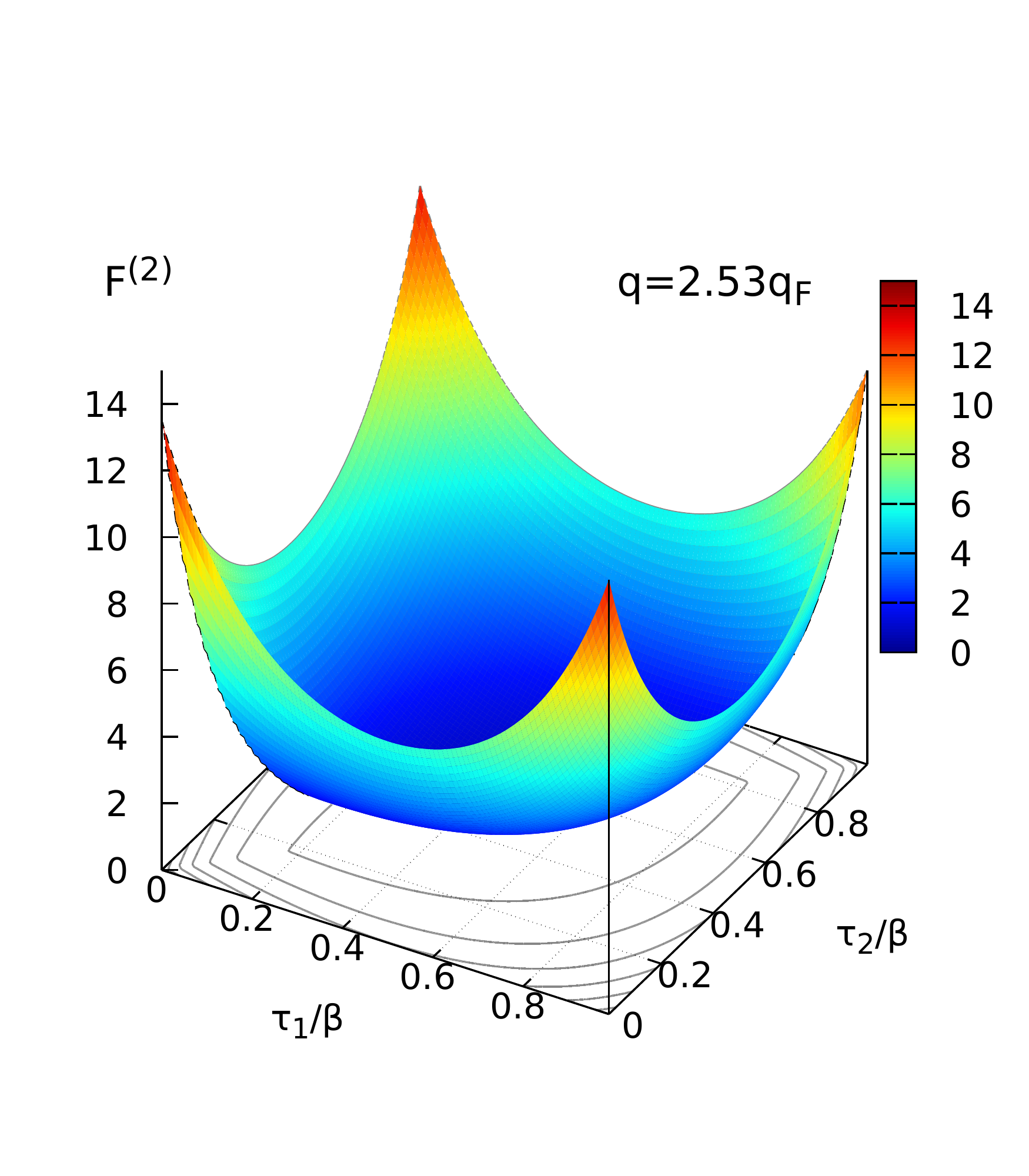}\includegraphics[width=0.485\textwidth]{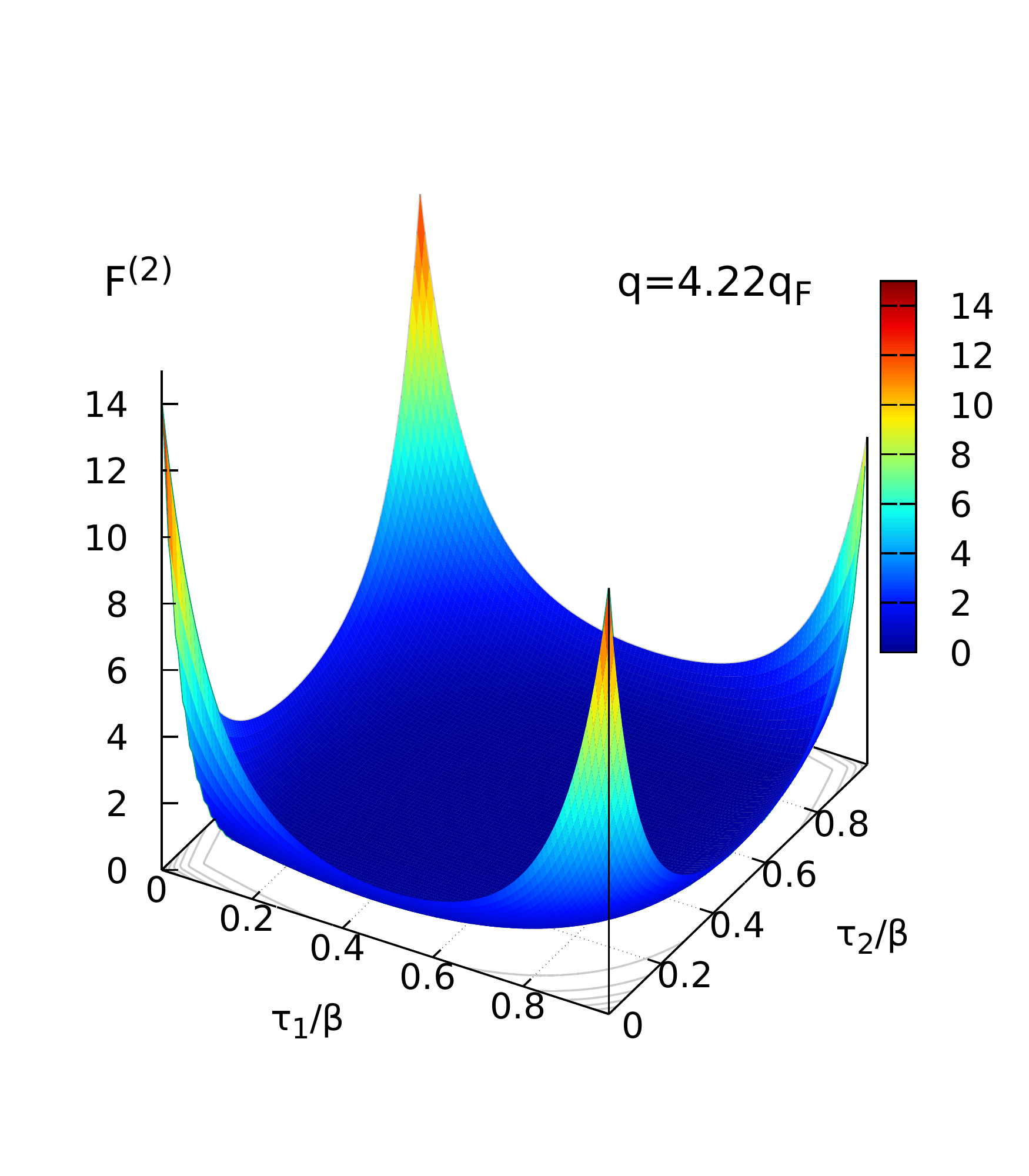}\\\vspace*{-1.cm}
\caption{\label{fig:F2_rs2_theta2} Imaginary-time response function $F^{(2)}(q,\tau_1,\tau_2)$ [cf.~Eq.~(\ref{eq:F2})] for different wave numbers $q$ in the $\tau_1$-$\tau_2$-plane for the unpolarized UEG with $N=14$ at $r_s=2$ and $\theta=2$.
}
\end{figure*} 

Let us start our investigation of the quadratic density response of the UEG by looking at the basic quantity, i.e., the quadratic ITCF $F^{(2)}(\mathbf{q},\tau_1,\tau_2)$ defined in Eq.~(\ref{eq:F2}). In Fig.~\ref{fig:F2_rs2_theta2}, we show our new PIMC results for this function in the $\tau_1$-$\tau_2$-plane for four different values of $q=|\mathbf{q}|$ for $N=14$ unpolarized electrons at $r_s=2$ and $\theta=2$. 
This value of the density parameter is of high relevance for contemporary research in the field of warm dense matter and can be probed, for example, in experiments with aluminum~\cite{Ramakrishna_PRB_2021,Sperling_PRL_2015}. Furthermore, we note that we depict $F^{(2)}(\mathbf{q},\tau_1,\tau_2)$ on the same colour- and $z$-scale for all panels to allow a direct comparison between the results for the different wave numbers. 

The top left panel corresponds to $q\approx0.84q_\textnormal{F}$, with
\begin{eqnarray}
q_\textnormal{F} = \left(
\frac{9\pi}{4}
\right)^{1/3} \frac{1}{r_s}
\end{eqnarray}
being the Fermi wave number~\cite{quantum_theory}, which corresponds to $\mathbf{q}=(2\pi/L,0,0)^T$ and, thus, constitutes the smallest possible value of $q$ that can be realized in this finite simulation cell. First and foremost, we find that $F^{(2)}(\mathbf{q},\tau_1,\tau_2)$ is symmetric around the diagonals defined by the relations $\tau_1=\tau_2$ and $\tau_1=\beta-\tau_2$, which can be seen particularly well by looking at the contour lines at the bottom. Secondly, the ITCF exhibits a very smooth behaviour in the entire $\tau_1$-$\tau_2$-plane, which is of great practical benefit for the numerical evaluation of Eq.~(\ref{eq:chi2_int}) discussed below. Upon increasing $q$ by a factor of two (top right), the slope towards the minimum at $\tau_1=\tau_2=\beta/2$ becomes substantially steeper. This trend is further exacerbated both for $q=2.53q_\textnormal{F}$ (bottom left) and $q=4.22q_\textnormal{F}$ (bottom right). In particular, $F^{(2)}(\mathbf{q},\tau_1,\tau_2)$ exhibits a nearly flat plateau around the center in the latter case, and most information about the density correlations is encoded in the steep slopes around small/large values of both time arguments. The reason for this behaviour can directly be seen from Eq.~(\ref{eq:triple_reconstruction}): the intricate correlation structure that is encoded in the triple dynamic structure factor $S(\mathbf{q},\omega_1,\mathbf{q},\omega_2)$ is exponentially damped by the kernel function $e^{-\omega_1\tau_1-\omega_2\tau_2}$ for large $\tau_1$ and/or $\tau_2$.
From a practical point of view, this implies that the numerical evaluation of Eq.~(\ref{eq:chi2_int}) will require an increasing number of imaginary-time points $P$ for large $q$, which, however, does not pose a fundamental obstacle.

\begin{figure}\centering
\vspace*{-1.5cm}\includegraphics[width=0.485\textwidth]{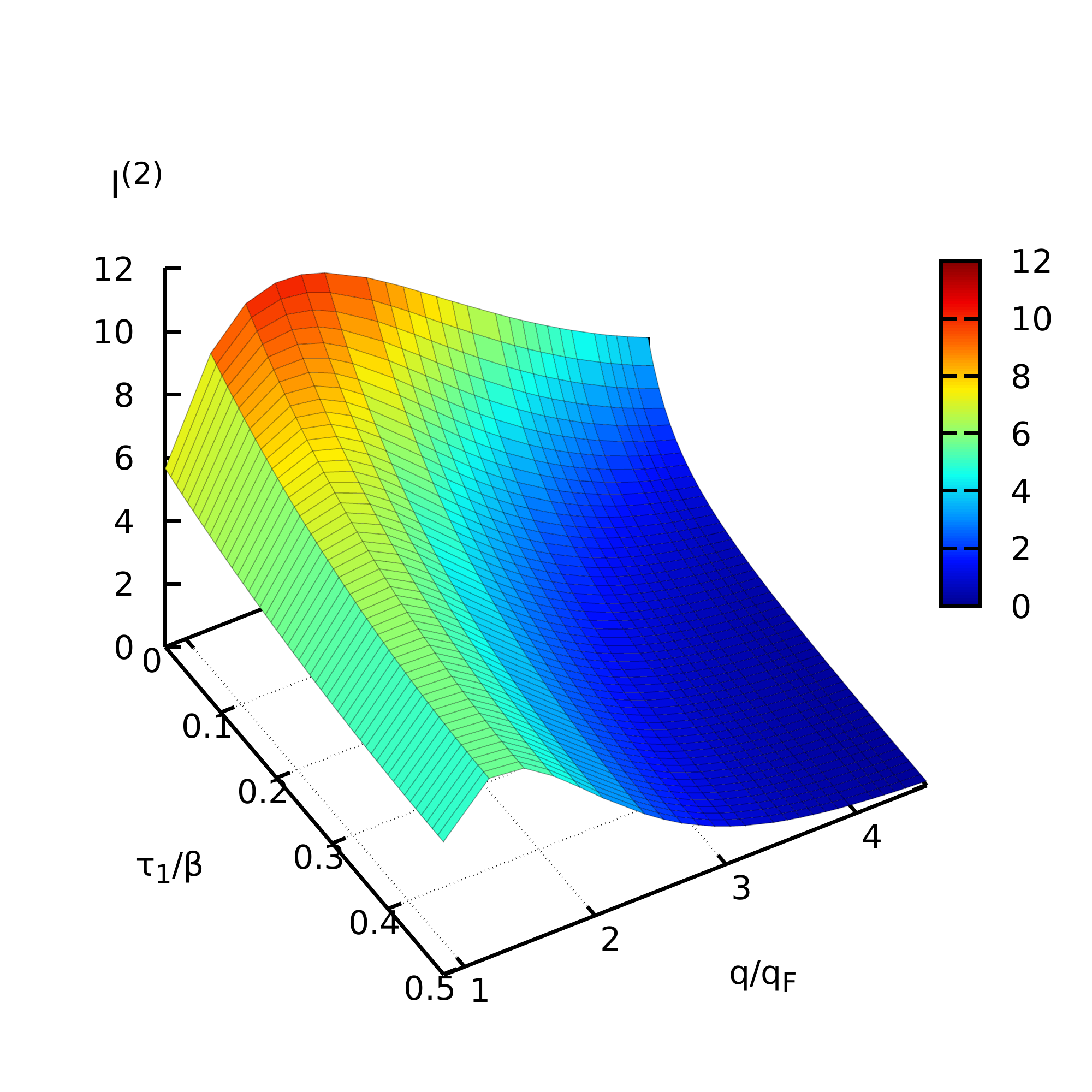}\\\vspace*{-1.5cm}\includegraphics[width=0.485\textwidth]{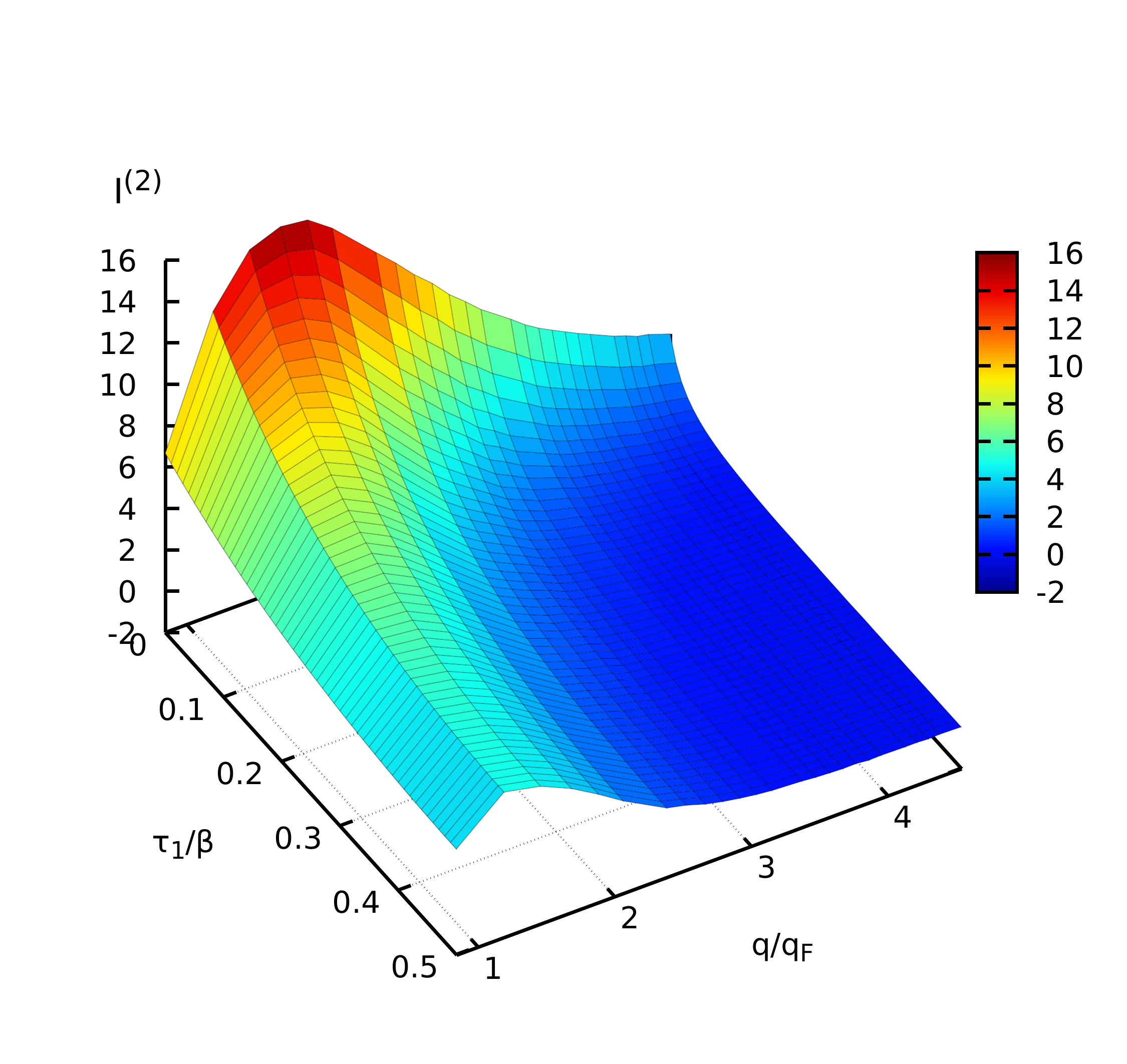}
\caption{\label{fig:I2_rs2} Imaginary-time response function $I{(2)}(q,\tau_1)$ [cf.~Eq.~(\ref{eq:I2})] for the unpolarized UEG with $N=14$ at $r_s=2$ and $\theta=2$ (top), $\theta=1$ (bottom).
}
\end{figure}

\begin{figure}\centering
\includegraphics[width=0.485\textwidth]{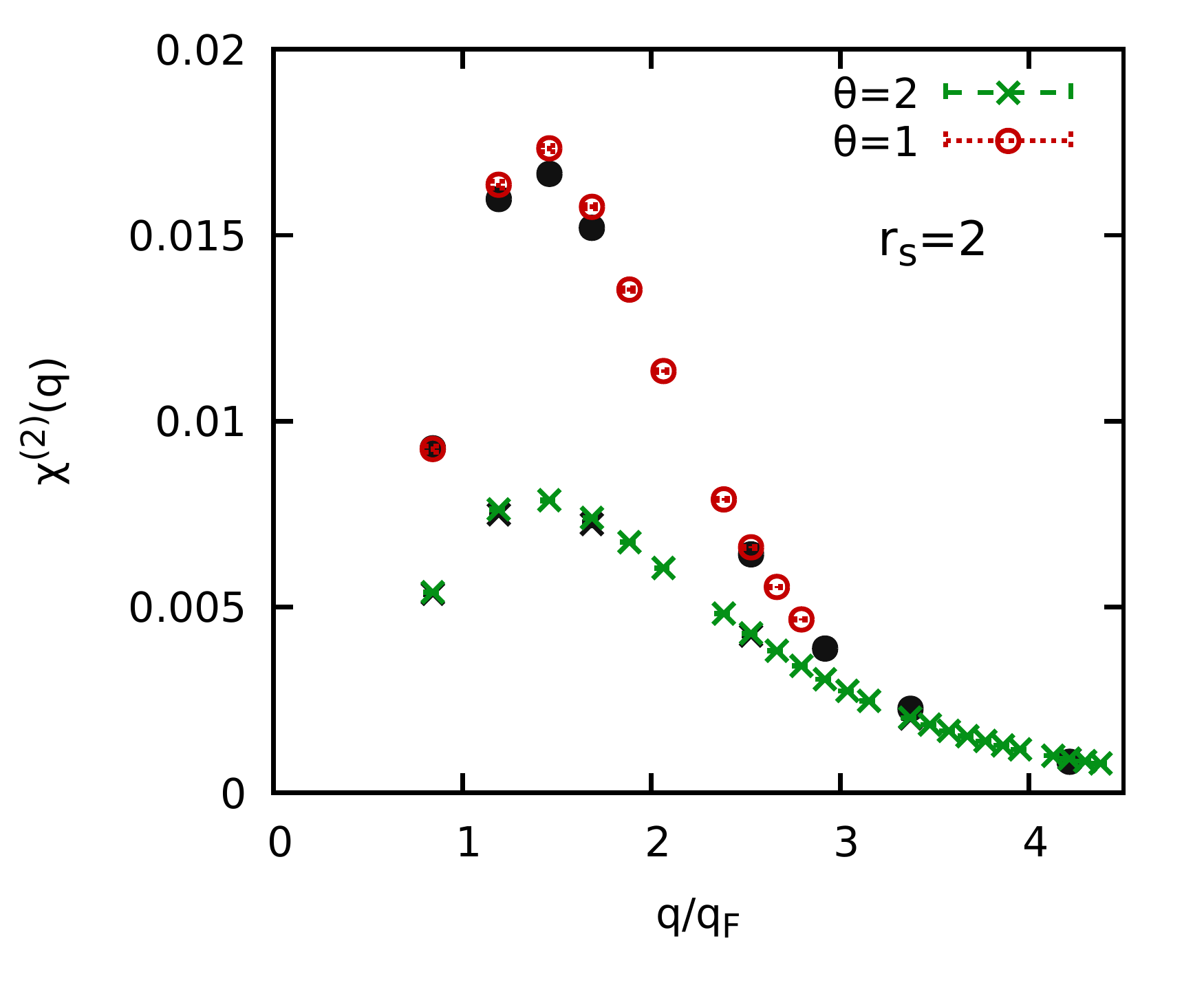}
\caption{\label{fig:quadratic_rs2} PIMC results for the quadratic density response function at the second harmonic, $\chi^{(2)}(\mathbf{q})$, for $N=14$ unpolarized electrons at $r_s=2$ with $\theta=2$ (green crosses) and $\theta=1$ (and red circles). The coloured symbols depict to our new data [cf.~Eq.~(\ref{eq:chi2_int})], and the corresponding black symbols the direct PIMC results [cf.~Eq.~(\ref{eq:n_second})] taken from Ref.~\cite{Dornheim_PRR_2021}.
}
\end{figure}

Let us proceed with our analysis by investigating the reduced quadratic ITCF $I^{(2)}(\mathbf{q},\tau_1)$, which in shown in Fig.~\ref{fig:I2_rs2} in the entire relevant $\tau_1$-$q$-plane. As this function is symmetric around $\tau_1=\beta/2$, i.e., $I^{(2)}(\mathbf{q},\tau_1)$ = $I^{(2)}(\mathbf{q},\beta-\tau_1)$, it is sufficient to restrict ourselves to $\tau_1\in[0,\beta/2]$.
The top panel has been obtained for $\theta=2$, i.e., for the same conditions as in Fig.~\ref{fig:F2_rs2_theta2}. Evidently, $I^{(2)}(\mathbf{q},\tau_1)$, too, is fairly smooth over the entire parameter-range, which further illustrates the comparably small requirements regarding the number of imaginary-time grid points for the numerical integration to compute $\chi^{(2)}(\mathbf{q})$.
In addition, we find a maximum around $q\approx1.5q_\textnormal{F}$ for all values of $\tau_1$. This is most likely an electronic exchange--correlation effect, and a similar behaviour has been observed for the IT intermediate scattering function $F(\mathbf{q},\tau)$, see e.g.~Refs.~\cite{dornheim_dynamic,dornheim_electron_liquid,dynamic_folgepaper}. For large wave numbers $q$, the slope along the $\tau$-direction becomes steeper, which is consistent to the trends observed in the full ITCF $F^{(2)}(\mathbf{q},\tau_1,\tau_2)$ shown in Fig.~\ref{fig:F2_rs2_theta2} above.

The bottom panel of Fig.~\ref{fig:I2_rs2} shows the same information, but obtained for half the temperature, $\theta=1$. While the overall behaviour is similar to the top panel, the corresponding features like the correlation-induced maximum at intermediate wave numbers are significantly more pronounced due to the increased coupling strength.

\begin{figure}\centering
\includegraphics[width=0.485\textwidth]{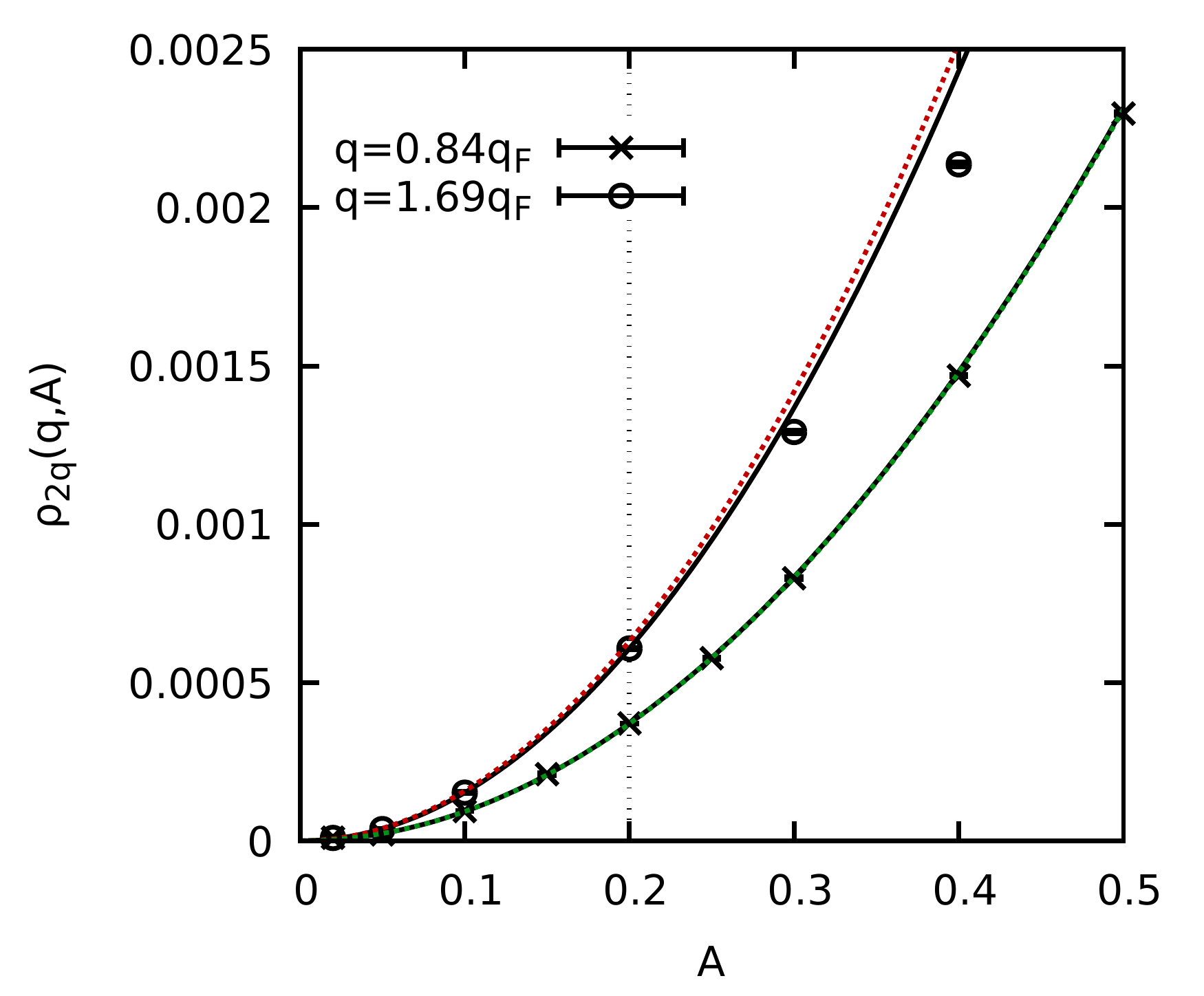}
\caption{\label{fig:A_rs2}Dependence of the density response at the second harmonic $\mathbf{k}=2\mathbf{q}$ [cf.~LHS.~of Eq.~(\ref{eq:n_second})] on the perturbation amplitude $A$ for $N=14$ unpolarized electrons at $r_s=2$ and $\theta=1$. Black crosses (circles): PIMC data for $q=0.84q_\textnormal{F}$ ($q=1.69q_\textnormal{F}$) taken from Ref.~\cite{Dornheim_PRR_2021}; black lines: quadratic fits, RHS.~of Eq.~(\ref{eq:n_second}). Coloured dotted lines: new ITCF-based results using $\chi^{(2)}(\mathbf{q})$ computed from Eq.~(\ref{eq:chi2_int}).
}
\end{figure}

\begin{figure*}\centering
\vspace*{-1.25cm}\includegraphics[width=0.485\textwidth]{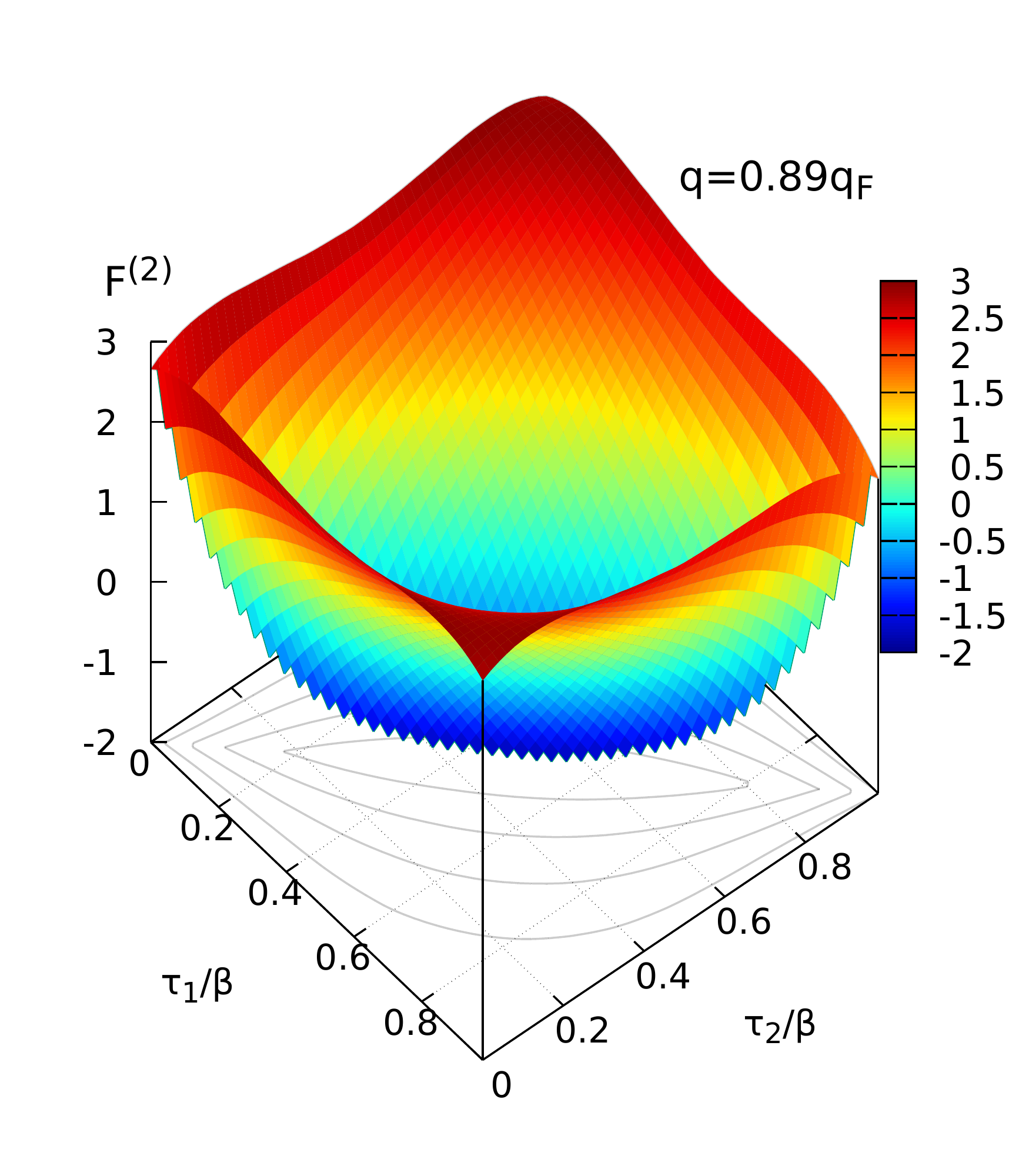}\includegraphics[width=0.485\textwidth]{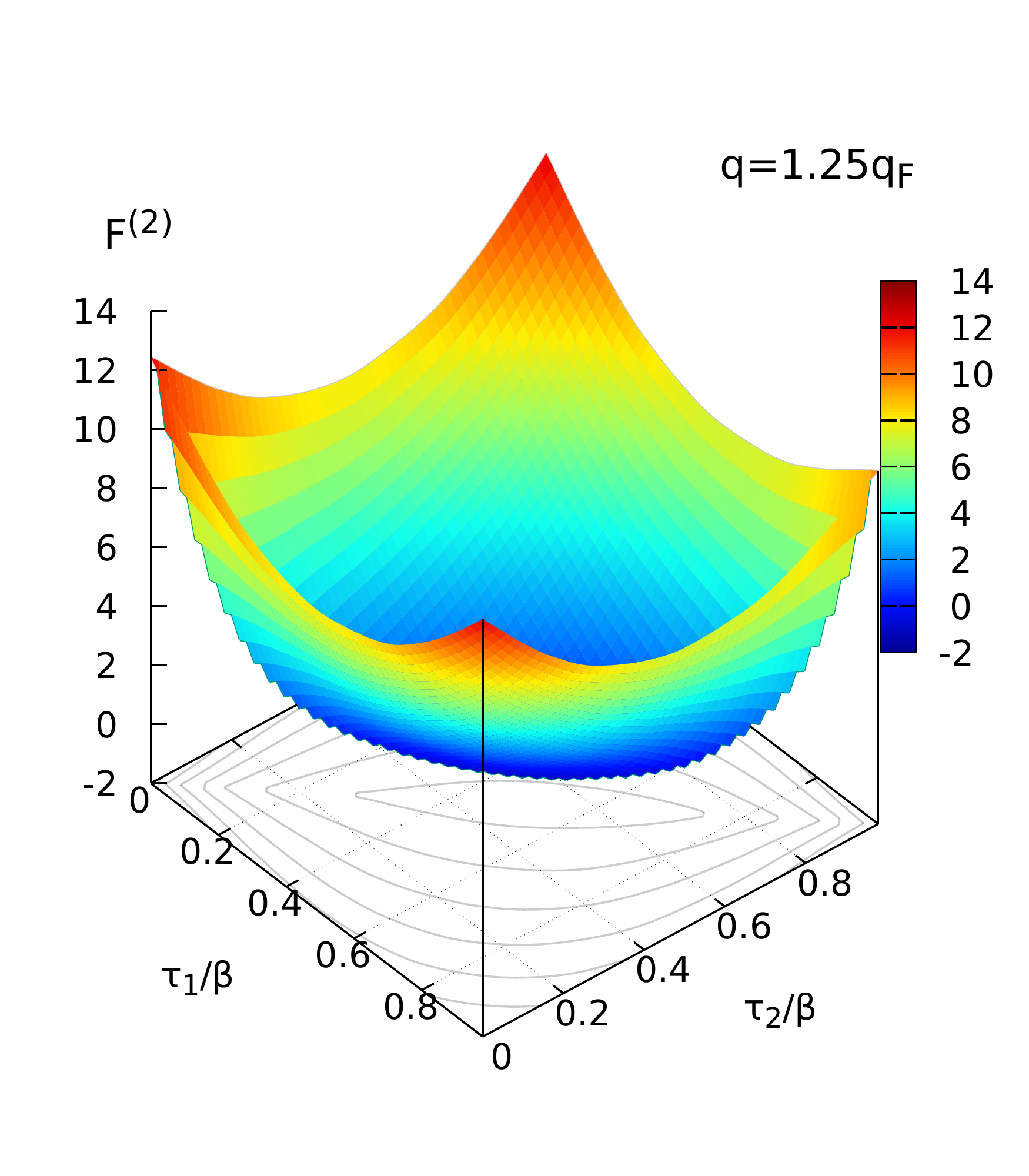}\vspace*{-0.7cm}
\caption{\label{fig:F2_rs6}Imaginary-time response function $F^{(2)}(q,\tau_1,\tau_2)$ [cf.~Eq.~(\ref{eq:F2})] for different wave numbers $q$ in the $\tau_1$-$\tau_2$-plane for the unpolarized UEG with $N=34$ at $r_s=6$ and $\theta=1$.
}
\end{figure*}

Let us next consider Fig.~\ref{fig:quadratic_rs2}, where we analyze the quadratic response function at the second harmonic of the original perturbation, $\chi^{(2)}(\mathbf{q})$. In particular, the green crosses ($\theta=2$) and red circles ($\theta=1$) have been obtained from the numerical integration along the $\tau_1$- and $\tau_2$-directions [cf.~Eq.~(\ref{eq:chi2_int})]. In addition, the corresponding dark symbols have been computed from direct PIMC simulations~\cite{Dornheim_PRR_2021} of the harmonically perturbed system by performing quadratic fits to the density response at the second harmonic according to Eq.~(\ref{eq:n_second}).
First and foremost, we note the excellent agreement between these independent data sets, which constitutes a strong empirical confirmation of our new approach. Moreover, we again stress that both the red and green data sets have been obtained from a single PIMC simulation of the unperturbed UEG, whereas the computation of the (much sparser) black data sets has taken more than $50$ times the computational effort for each temperature. The only small yet significant deviations between the coloured and dark symbols appear in the three data points around $q=1.5q_\textnormal{F}$ for $\theta=1$.

To understand the origin of this finding, we show the actual dependence of the density response at the second harmonic of the original perturbation, $\mathbf{k}=2\mathbf{q}$, in Fig.~\ref{fig:A_rs2}. More specifically, the black crosses and circles correspond to the PIMC expectation values for the LHS. of Eq.~(\ref{eq:n_second}) for $q=0.84q_\textnormal{F}$ and $q=1.69q_\textnormal{F}$, respectively, and have been taken from Ref.~\cite{Dornheim_PRR_2021}. Further, the corresponding solid black lines depict quadratic fits, cf.~the RHS.~of Eq.~(\ref{eq:n_second}) taking into account data points for $A\in[0,0.2]$ (vertical dotted grey line). Finally, the dotted green and red lines have been obtained using our new results for $\chi^{(2)}(\mathbf{q})$ that have been obtained by numerically evaluating $F^{(2)}(\mathbf{q},\tau_1,\tau_2)$. For the smaller wave number, the parabolas from the direct fit and from the ITCF are in perfect agreement with each other, and reproduce the PIMC data points well even outside of the given fitting interval. For the larger value of $q$, however, the parabola computed from the ITCF noticeably exceeds the black curve, which has been reflected in the deviation observed in Fig.~\ref{fig:quadratic_rs2}. Evidently, both the black and red curve nicely reproduce the PIMC data points for $A\leq0.1$ within the given Monte Carlo error bars. The deviation between the two curves is, thus, caused by the single data point at $A=0.2$, which has been included into the quadratic fit in Ref.~\cite{Dornheim_PRR_2021} and deviates by about $4\%$ from the ITCF-based result. Yet, this shift appears to be spurious, as the PIMC data points for the actual induced density at $\mathbf{k}=2\mathbf{q}$ increasingly deviate from both parabolas for $A>0.2$. The most likely explanation for the observed deviation between the direct and ITCF-based estimations of $\chi^{(2)}(\mathbf{q})$ is therefore given by the truncation of Eq.~(\ref{eq:n_second}) at the quadratic level. In practice, this means that the fit will result in coefficients that are systematically too low, as they have to accommodate higher-order contributions to the actual PIMC data that have the opposite sign of $\chi^{(2)}(\mathbf{q})$. On the other hand, this further highlights the value of our new approach, as it will yield, by definition \emph{clean} results for the respective nonlinear response function without any spurious contributions from other terms that are of a different order in $A$.

\begin{figure}\centering
\vspace*{-1.5cm}\includegraphics[width=0.485\textwidth]{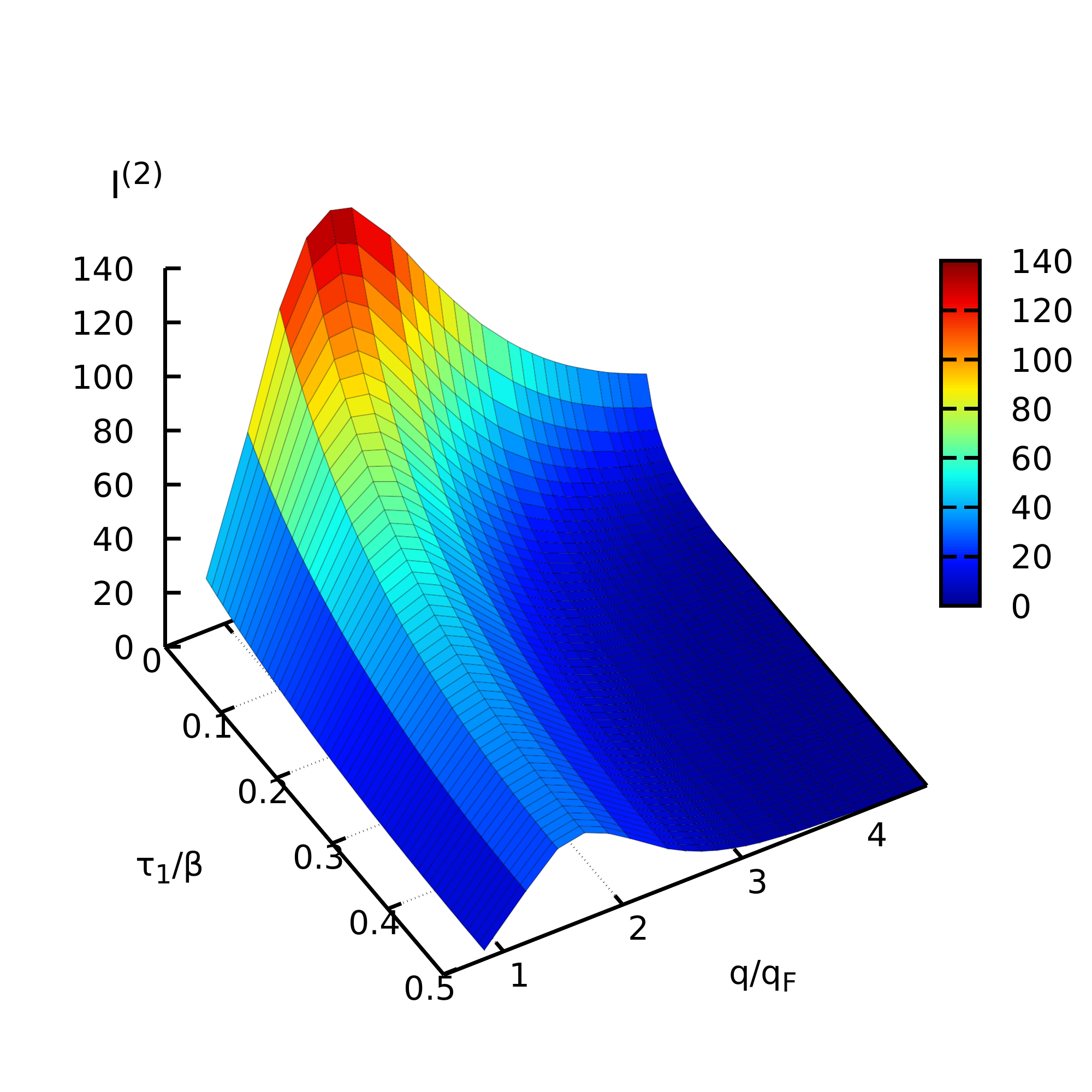}
\caption{\label{fig:I2_rs6_theta1} Imaginary-time response function $I{(2)}(q,\tau_1)$ [cf.~Eq.~(\ref{eq:I2})] for the unpolarized UEG with $N=14$ at $r_s=6$ and $\theta=1$.
}
\end{figure} 

From a physical perspective, we note that the data for $\chi^{(2)}(\mathbf{q})$ shown in Fig.~\ref{fig:quadratic_rs2}, too, exhibits a maximum around $q\approx1.5q_\textnormal{F}$, which can be understood in the following way~\cite{Dornheim_PRR_2021}: in the limit of small wave numbers $q\to0$, the UEG is perfectly screened~\cite{kugler_bounds}, and, correspondingly, does not respond to an external perturbation. For large wave numbers, on the other hand, collective effects play an increasingly diminished role and the density response will eventually be dominated by single particle effects that are comparably small in magnitude. We note that the same trends also universally apply for the linear response of the UEG, see e.g. Refs.~\cite{quantum_theory,dornheim_HEDP}.

The second physical trend that becomes evident from Fig.~\ref{fig:quadratic_rs2} is the comparably larger magnitude of the quadratic density response for the lower temperature, $\theta=1$. Again, this behaviour can be traced back to the importance of collective effects, which decreases with increasing temperature.

Let us next repeat this analysis at a lower density, $r_s=6$. These conditions can be realized in evaporation experiments~\cite{karasiev_importance,low_density1,low_density2} and constitute an ideal test bed for the study of electronic exchange--correlation effects. In Fig.~\ref{fig:F2_rs6}, we show the quadratic ITCF $F^{(2)}(q,\tau_1,\tau_2)$ in the $\tau_1$-$\tau_2$-plane for two relevant wave numbers $q$. More specifically, the left panel corresponds to $\mathbf{q}=2\pi/L(1,1,0)^T$ ($q\approx0.89q_\textnormal{F}$) and exhibits a rich, non-monotonous structure that was absent for the case of $r_s=2$ studied earlier (see Fig.~\ref{fig:F2_rs2_theta2} above) for a similar wave number $q$. Therefore, it is safe to conclude that these nontrivial features constitute a direct consequence of the increased coupling strength between the electrons at this lower density. 
The right panel corresponds to $q\approx1.25q_\textnormal{F}$ where $F^{(2)}(q,\tau_1,\tau_2)$ is overall increased in magnitude, but exhibits a less rich structure.

As the next step, we show the full $\tau_1$-$q$-dependence of the corresponding reduced ITCF $I^{(2)}(q,\tau_1)$ in Fig.~\ref{fig:I2_rs6_theta1}. While the general behaviour is again similar to the case of $r_s=2$ (cf.~Fig.~\ref{fig:I2_rs2}), the reduced ITCF exhibits a more pronounced maximum around $q\approx1.5q_\textnormal{F}$. Indeed, this is consistent with our earlier interpretation of this feature as an electronic exchange--correlation effect. Furthermore, we find a monotonous decay of $I^{(2)}(q,\tau_1)$ (until $\tau=\beta/2$) for all wave numbers $q$.

\begin{figure}\centering
\includegraphics[width=0.485\textwidth]{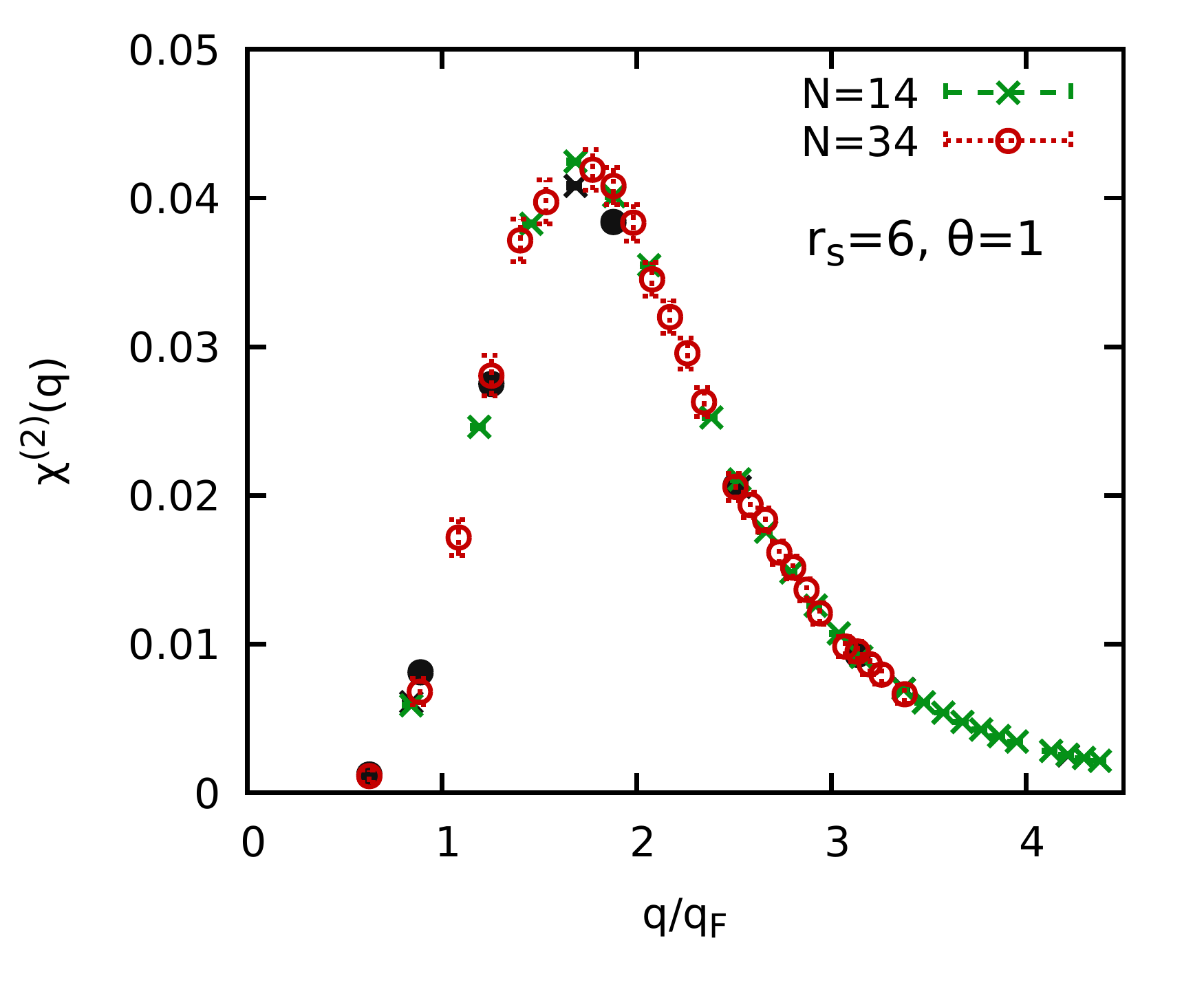}
\caption{\label{fig:quadratic_rs6} PIMC results for the quadratic density response function at the second harmonic, $\chi^{(2)}(\mathbf{q})$, for $N=14$ (green crosses) and $N=34$ (red circles) unpolarized electrons at $r_s=6$ and $\theta=1$. The coloured symbols depict to our new data [cf.~Eq.~(\ref{eq:chi2_int})], and the corresponding black symbols the direct PIMC results [cf.~Eq.~(\ref{eq:n_second})] taken from Ref.~\cite{Dornheim_PRR_2021}.
}
\end{figure}

Let us conclude our analysis of the quadratic density response of the UEG by considering $\chi^{(2)}(\mathbf{q})$ for the lower density, $r_s=6$. The results are shown in Fig.~\ref{fig:quadratic_rs6}, where the red circles and green crosses have been obtained by numerically evaluating the double integral in Eq.~(\ref{eq:chi2_int}) for $N=34$ and $N=14$ electrons, respectively. As it is expected, they are in excellent agreement with the corresponding black points that have been obtained on the basis of the perturbed electron gas in Ref.~\cite{Dornheim_PRR_2021}.
For $N=34$, it is even possible to access the small wave-number regime where $\chi^{(2)}(\mathbf{q})$ has decayed by approximately two orders of magnitude compared to the maximum value. At the same time, we confirm the previous findings~\cite{Dornheim_PRL_2020,Dornheim_PRR_2021} that finite-size effects do not play an important role for these parameters.

\subsection{Cubic density response\label{sec:cubic_results}}

\begin{figure}\centering
\includegraphics[width=0.485\textwidth]{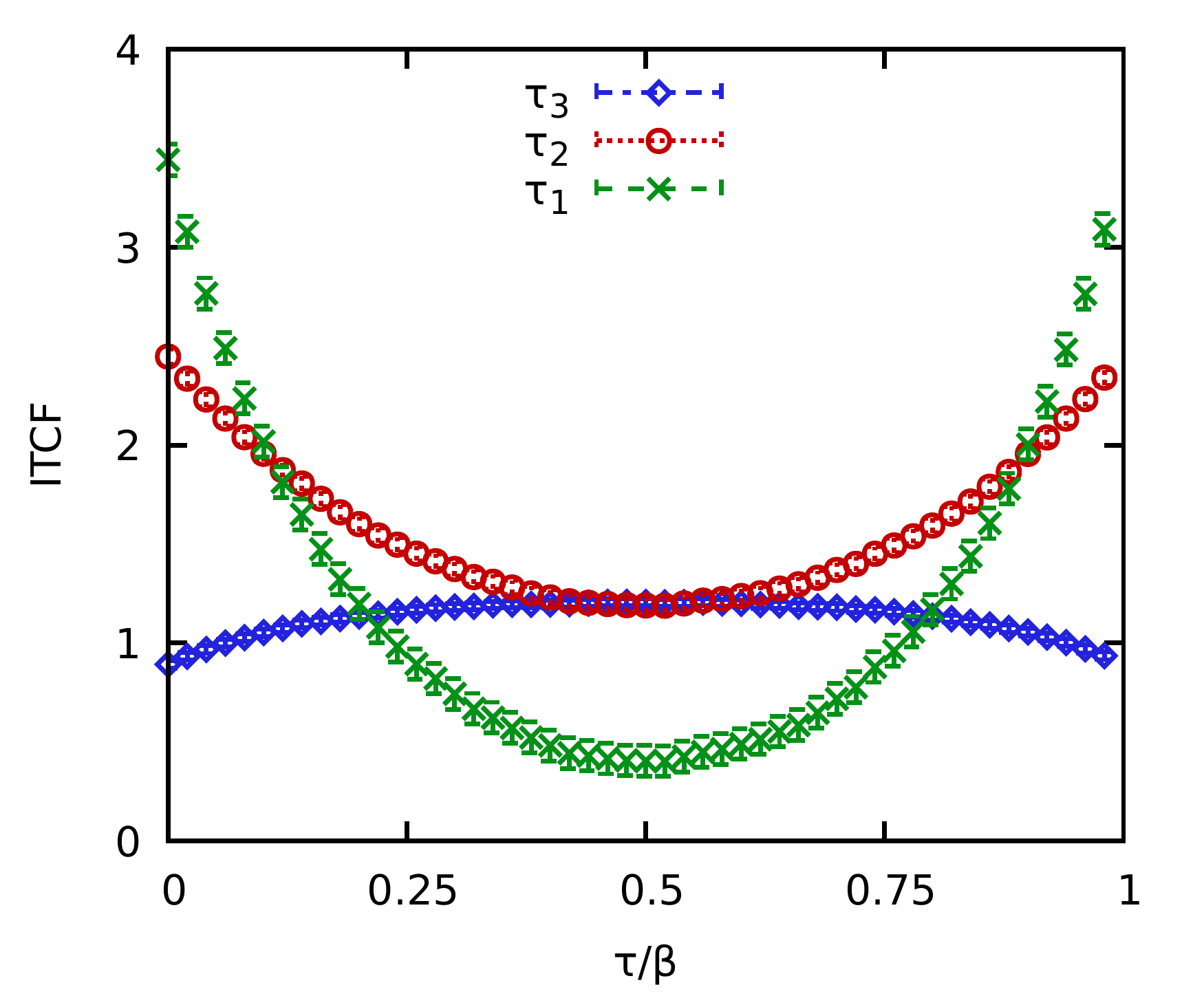}
\caption{\label{fig:analyze_ITCF} Illustration of different integration levels of $F^{(3)}(\mathbf{q},\tau_1,\tau_2,\tau_3)$ [cf.~Eq.~(\ref{eq:F3})] for $N=14$, $r_s=2$, and $\theta=1$, for $\mathbf{q}=(2\pi/L,0,0)^T$. Blue diamonds: $F^{(3)}(\mathbf{q},0,0,\tau_3)$; red circles: $I^{(3)}(\mathbf{q},0,\tau_2)$ [Eq.~(\ref{eq:I3})]; green crosses: $J^{(3)}(\mathbf{q},\tau_1)$ [Eq.~(\ref{eq:J3})].
}
\end{figure}

Let us next explore the relation between the imaginary-time path-integral structure of the system and the cubic density response. As a representative example, we consider the cubic density response at the third harmonic, cf.~Eq.~(\ref{eq:n_third}) above. 
Since the graphical depiction of the corresponding four-dimensional ITCF $F^{(3)}(\mathbf{q},\tau_1,\tau_2,\tau_3)$
is fairly challenging, we here restrict ourselves to showing its dependency along the distinct directions in $\tau$-space. The results are shown in Fig.~\ref{fig:analyze_ITCF} for $r_s=2$ and $\theta=1$.

More specifically, the blue diamonds show our new PIMC data for $F^{(3)}(\mathbf{q},0,0,\tau_3)$, i.e., along the $\tau_3$ direction. While being symmetric around $\tau=\beta/2$, as it is expected, the data set exhibits an unusual curvature with a maximum in the center. 
Averaging along the $\tau_3$-direction gives us the reduced function $I^{(3)}(\mathbf{q},\tau_1,\tau_2)$ [cf.~Eq.~(\ref{eq:I3})], and we show $I^{(3)}(\mathbf{q},0,\tau_2)$ along the $\tau_2$-direction as the red circles in Fig.~\ref{fig:analyze_ITCF}. Evidently, the reduced ITCF is similarly smooth as $F^{(3)}$, although the curvature is flipped, with the familiar minimum at $\tau=\beta/2$.
The green crosses have been obtained for one additional integration along the $\tau_2$-direction and show our results for $J^{(3)}(\mathbf{q},\tau_1)$ [cf.~Eq.~(\ref{eq:J3})].

\begin{figure}\centering
\includegraphics[width=0.485\textwidth]{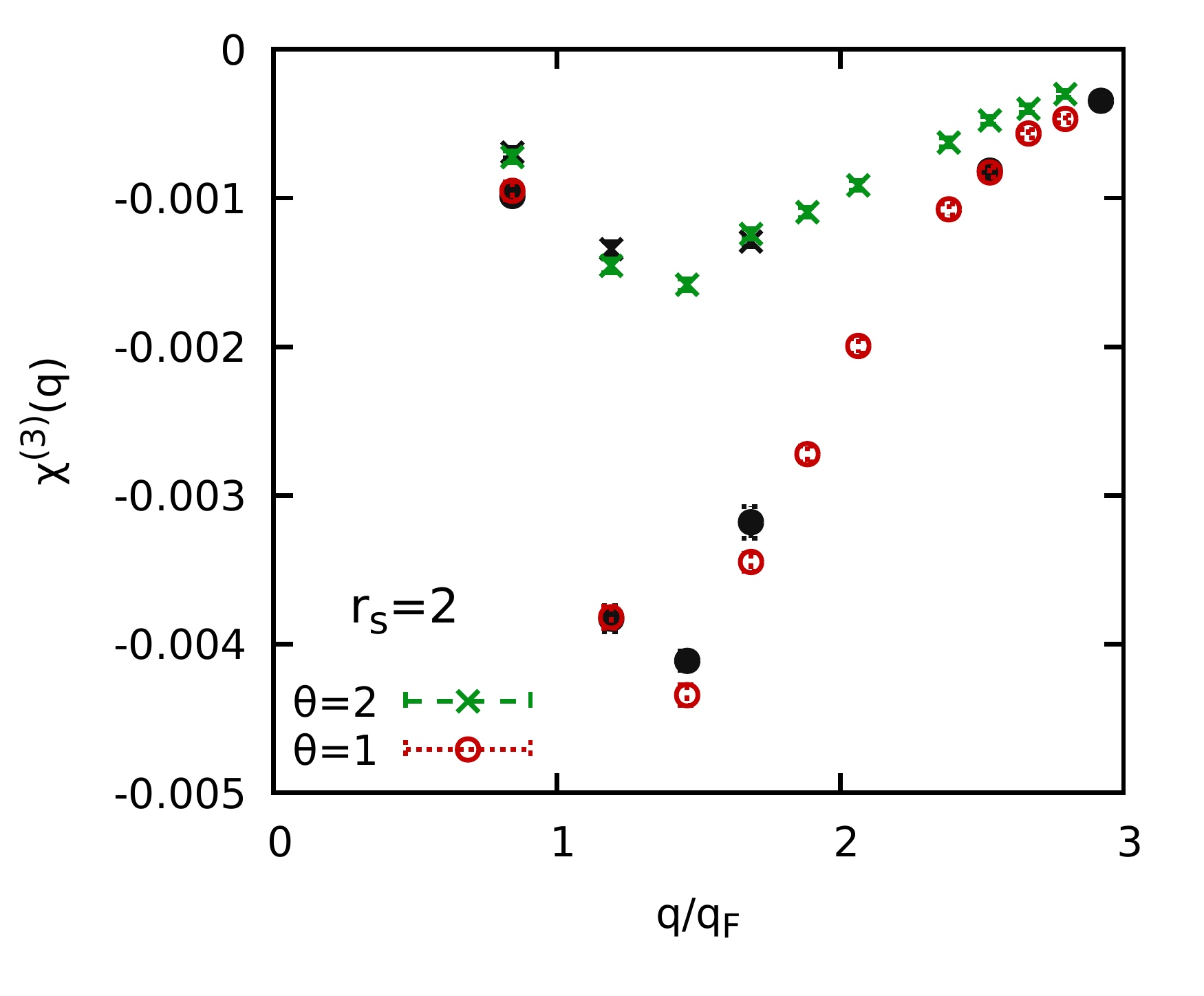}\\
\includegraphics[width=0.485\textwidth]{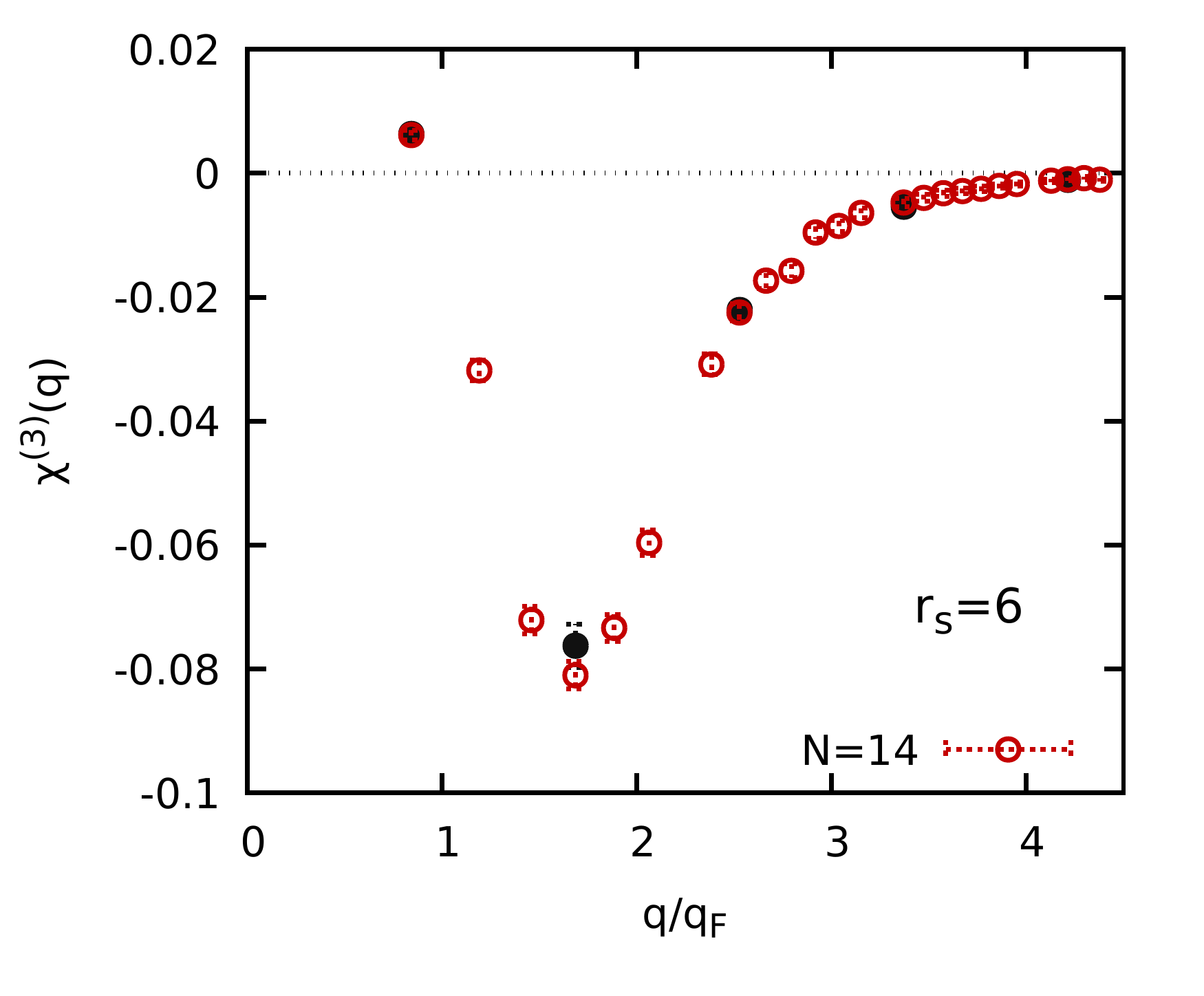}
\caption{\label{fig:third} Cubic response function of the third harmonic $\chi^{(3)}(\mathbf{q})$ of the unpolarized UEG, cf.~Eq.~(\ref{eq:chi3_int}). top: $N=14$, $r_s=2$ with the green crosses and red circles showing results for $\theta=2$ and $\theta=1$; bottom: $r_s=6$, $\theta=1$ with the red circles showing results for $N=14$.
The black symbols show the corresponding results obtained from PIMC simulations of the harmonically perturbed UEG and have been taken from Ref.~\cite{Dornheim_PRR_2021}. 
}
\end{figure}

Finally, we show our results for $\chi^{(3)}(\mathbf{q})$ that have been obtained by numerically evaluating Eq.~(\ref{eq:chi3_int}) in Fig.~\ref{fig:third}. The top panel shows results for $r_s=2$, with the green crosses and red circles depicting results for $\theta=2$ and $\theta=1$, respectively. In addition, the corresponding black symbols have been obtained on the basis of PIMC simulations of the harmonically perturbed UEG [cf.~Eq.~(\ref{eq:n_third})] and are in perfect agreement to our new results based on the ITCF $F^{(3)}(\mathbf{q},\tau_1,\tau_2,\tau_3)$.

The bottom panel shows a similar study for $r_s=6$ and $\theta=1$, with the same excellent agreement between the two independent data sets. Interestingly, $\chi^{(3)}(\mathbf{q})$ changes its sign for the smallest wave number. Yet, this constitutes a real physical effect and is indeed correctly reflected in both the red and black data point.

Let us conclude our analysis of the PIMC approach to the nonlinear density response based on novel ITCFs with a practical remark. Similarly to the quadratic density response function $\chi^{(2)}(\mathbf{q})$ discussed above, the numerical integration of $F^{(3)}(\mathbf{q},\tau_1,\tau_2,\tau_3)$ is well behaved and a small number of imaginary-time samples $P\sim50-100$ is sufficient. This only changes for larger wave numbers $q$, when the slope along the respective $\tau$-directions becomes steeper. 
In addition, we mention that the estimation of $F^{(3)}(\mathbf{q},\tau_1,\tau_2,\tau_3)$ leads to a noticeable slowing down of our PIMC simulation in practice, as it requires three nested loops over the $P$ slices. Yet, this does not negate the benefits of our method, as we can still obtain the full $\mathbf{q}$-dependence from a single simulation of the unperturbed system.

\section{Summary and Outlook\label{sec:summary}}

In this work, we have presented new relations between the nonlinear density response of a system in terms of density correlations that are encoded in generalized ITCFs. While previous PIMC studies of nonlinear effects~\cite{Dornheim_PRL_2020,Dornheim_PRR_2021} were based on multiple simulations for different perturbation amplitudes $A$ \emph{for each individual wave number}, our new scheme allows one to obtain the complete nonlinear density response over the entire $\mathbf{q}$-range of interest simply from a single PIMC simulation of the unperturbed system. This leads to a typical reduction of the computational effort by two orders of magnitude, and thus facilitates the realization of extensive parameter scans.

As a practical example of high relevance, we have demonstrated our approach for the uniform electron gas at warm dense matter conditions. Specifically, we have presented the first results for the generalized higher-order ITCFs $F^{(2)}(\mathbf{q},\tau_1,\tau_2)$ and $F^{(3)}(\mathbf{q},\tau_1,\tau_2,\tau_3)$ for different densities and temperatures. In addition, the subsequent numerical integration of these data has given us direct access to the quadratic response function of the second harmonic of the original perturbation $\chi^{(2)}(\mathbf{q})$ and to the cubic response function of the third harmonic $\chi^{(3)}(\mathbf{q})$. The comparison to previous results based on extensive PIMC simulations of the harmonically perturbed electron gas has revealed an excellent agreement between these independent data sets and, thus, constitutes a strong empirical validation of our new approach.

Lastly, we have also given a relation between the density--density--density ITCF $\mathscr{F}(\mathbf{q}_a,\tau_a;\mathbf{q}_b,\tau_b;\mathbf{q}_c,\tau_c)$
and the triple dynamic structure factor $S(\qv_1,\omega_1;\qv_2,\omega_2)$, which constitutes the basis for a new type of analytic continuation.

We are confident that our results will open up new avenues for exciting research in many areas of statistical physics, quantum chemistry, and related disciplines. First and foremost, our scheme gives one direct access to the nonlinear density response of all systems that can be simulated with the PIMC method. This includes such diverse applications as quantum-dipole system~\cite{Filinov_PRA_2012,Filinov_PRA_2016,Dornheim_PRA_2020}, ultracold atoms like $^4$He~\cite{cep,Boninsegni_MaxEnt_revisited_2018}, confined nano clusters~\cite{Filinov_PRB_2008,mezza,Boninsegni_PRA_2013}, quantum crystals~\cite{PhysRevLett.76.4572,PhysRevLett.86.3851}, and exotic supersolids~\cite{Saccani_Supersolid_PRL_2012,Kora2019}. Of particular interest is so-called warm dense matter~\cite{new_POP,wdm_book}, for which nonlinear effects have already been shown to play an important role~\cite{Dornheim_PRL_2020,Dornheim_PRR_2021} and, in addition, might constitute a promising new method of diagnostics~\cite{moldabekov2021thermal}.

Furthermore, we mention that the computation of ITCFs is by no means limited to the PIMC method. For example, first results for $F(\mathbf{q},\tau)$ for the UEG in the ground state have been presented by Motta \emph{et al.}~\cite{Motta_JCP_2015} using the phaseless auxiliary field method. Other examples include variations of the widely used continuous-time quantum Monte Carlo method~\cite{PhysRevE.82.026701}, and the configuration PIMC approach~\cite{groth_prb_2016} at finite temperature.

Finally, the presented relation between $\mathscr{F}(\mathbf{q}_a,\tau_a;\mathbf{q}_b,\tau_b;\mathbf{q}_c,\tau_c)$
and $S(\qv_1,\omega_1;\qv_2,\omega_2)$ evokes the titillating possibility to study dynamic three-body correlations on an \emph{ab initio} level without any approximations on exchange--correlation effects.


\appendix
\section{Basic formulas\label{sec:derivations}}
The induced density is given according to a weak perturbation assumption~\cite{PhysRevB.37.9268,Bergara1999} by a power series in the external potential where the pre-factors are the linear  $\chi$, quadratic $\mathscr{Y}$, cubic $\mathscr{Z}$, etc. response functions
 \begin{align}
n_{\rm ind}(\vec r)\! &=\!\!\int\!\mathrm{d} \vec r^{\prime} \chi(\vec r, \vec r^{\prime}) V(\vec r^{\prime}) \nonumber\\
&+\int\! \mathrm{d} \vec r^{\prime} \mathrm{d} \vec r^{\prime \prime} \mathscr{Y}(\vec r,\vec r^{\prime},\vec r^{\prime \prime})V(\vec r^{\prime})V(\vec r^{\prime \prime}) \nonumber \\
&+\int\! \mathrm{d} \vec r^{\prime} \mathrm{d} \vec r^{\prime \prime} \mathrm{d} \vec r^{\prime \prime \prime} \mathscr{Z}(\vec r,\vec r^{\prime},\vec r^{\prime \prime}, \vec r^{\prime \prime \prime})V(\vec r^{\prime})V(\vec r^{\prime \prime})V(\vec r^{\prime  \prime \prime}) \nonumber\\
&+\cdots\ .
\label{eq:n_ind_re}
\end{align}

Taking into account that, for a homogeneous system, $\chi(\vec r, \vec r^{\prime})=\chi(\vec r- \vec r^{\prime})$, $\mathscr{Y}(\vec r,\vec r^{\prime},\vec r^{\prime \prime})= \mathscr{Y}(\vec r-\vec r^{\prime},\vec r-\vec r^{\prime \prime})$, and $\mathscr{Z}(\vec r,\vec r^{\prime},\vec r^{\prime \prime}, \vec r^{\prime \prime \prime})=\mathscr{Z}(\vec r-\vec r^{\prime}, \vec r-\vec r^{\prime \prime}, \vec r - \vec r^{\prime \prime \prime})$,  Eq.~(\ref{eq:n_ind_re}) can be rewritten in Fourier space as ($\Omega=L^3$ denotes the volume)
\begin{align}
n_{\rm ind}(\vec k)\! &=\! \chi(\vec k) V(\vec k)  \nonumber\\
& + \frac{1}{\Omega}\sum_{\vec k_2} \! \mathscr{Y}(\vec k-\vec k_2,\vec k_2)V(\vec k-\vec k_2)V(\vec k_2)\nonumber\\
&+\frac{1}{\Omega^2}\sum_{\vec k_2}\sum_{\vec k_3}\, \mathscr{Z}(\vec k-\vec k_2+\vec k_3,\vec k_2,\vec k_3) \nonumber\\
&\times V(\vec k-\vec k_2+\vec k_3)V(\vec k_2)V(\vec k_3) +\cdots,
\label{eq:n_ind_q}
\end{align}
where the notation $\vec k$ is used for the wave vector to avoid confusion with the wave-vector $\vec q$ of the external harmonic field, $V(\vec r)=2A\cos{(\vec q\cdot \vec r)}$. 

A single external harmonic perturbation at wave vector $\qv$ and amplitude $2A$ therefore leads to 
\begin{align}
n_{\rm ind}(\vec k)\! &=\! 2A\chi(\vec k) \delta(\vec k-\vec q)  \nonumber\\
& +4A^2 \mathscr{Y}(\vec k-\vec q,\vec q)
\delta(\vec k-2\vec q)\nonumber\\
&+6A^3\mathscr{Z}(\vec k-2\vec q,\vec q,\vec q)\delta(\vec k-3\vec q) \nonumber\\
&+18A^3\mathscr{Z}(\vec k-2\vec q,\vec q,\vec q)\delta(\vec k-\vec q) +\cdots.
\label{eq:n_ind_q}
\end{align}

\section*{Data Availability Statement}
The data that support the findings of this study are available from the corresponding author upon reasonable request.

\section*{Acknowledgments}
We gratefully acknowledge stimulating discussions with Michael Bonitz.

This work was partly funded by the Center for Advanced Systems Understanding (CASUS) which is financed by Germany's Federal Ministry of Education and Research (BMBF) and by the Saxon Ministry for Science, Culture and Tourism (SMWK) with tax funds on the basis of the budget approved by the Saxon State Parliament.
We gratefully acknowledge CPU-time at the Norddeutscher Verbund f\"ur Hoch- und H\"ochstleistungsrechnen (HLRN) under grant shp00026 and on a Bull Cluster at the Center for Information Services and High Performace Computing (ZIH) at Technische Universit\"at Dresden.

\bibliography{bibliography}

\begin{thebibliography}{103}
\bibitem [{bet()}]{beta_note}%
  \BibitemOpen
  \href@noop {} {}\bibinfo {note} {Strictly speaking, the \emph{imaginary time}
  is defined as $\tau\in[0,-i\beta]$. Yet, it has become the convention to
  instead use $\tau\to\tau/(-i)$, which we employ throughout this
  work.}\BibitemShut {Stop}%
\end{thebibliography}
\end{document}